\documentclass[onecolumn,tightenlines,showpacs,nofootinbib,preprintnumbers,
amsmath,amssymb]{revtex4}

\usepackage[usenames,dvipsnames]{color}

\usepackage{graphicx}
\usepackage{bm}
\usepackage{lscape}

\begin{document}


\title{Acquire information about neutrino parameters by detecting supernova neutrinos}

\author{Ming-Yang Huang$^{1}$\footnote{Email: hmy19151905@mail.bnu.edu.cn}, Xin-Heng Guo$^{1}$\footnote{Corresponding author, email: xhguo@bnu.edu.cn },
and Bing-Lin Young$^{2,3}$\footnote{Email: young@iastate.edu}}
\affiliation{\small $^{1}$College of Nuclear Science and
Technology, Beijing Normal University, Beijing 100875, China \\
\small $^2$ Department of Physics and Astronomy, Iowa State
University, Ames, Iowa 5001, USA \\
\small $^3$ Institute of Theoretical Physics, Chinese Academy of
Sciences, Beijing 100190, China}


\begin{abstract}
We consider the supernova shock effects, the
Mikheyev-Smirnov-Wolfenstein (MSW) effects, the collective effects,
and the Earth matter effects in the detection of type II supernova
neutrinos on the Earth. It is found that the event number of
supernova neutrinos depends on the neutrino mass hierarchy, the
neutrino mixing angle $\theta_{13}$, and neutrino masses. Therefore,
we propose possible methods to identify the mass hierarchy and
acquire information about $\theta_{13}$ and neutrino masses by
detecting supernova neutrinos. We apply these methods to some
current neutrino experiments.
\end{abstract}

\maketitle

\section{\label{intro}Introduction}

Supernovas (SNs) are extremely powerful explosions in the universe
which terminate the life of some stars \cite{Kotake}. They make the
catastrophic end of stars more massive than 8 solar masses, leaving
behind compact remnants such as neutron stars or black holes. The SN
explosion, such as SN1987A \cite{SN1987}, is one of the most
spectacular comic events and the source of new physics ideas
\cite{Dighe1}. Detecting SN neutrinos on the Earth has been a
subject of active investigation in astroparticle physics
\cite{Kotake}\cite{SN1987}\cite{Guo1}-\cite{Beacom1}. We may obtain
some information about the explosion mechanism of SN
\cite{Raffelt100} and the parameters of neutrinos
\cite{Guo1}\cite{Dasgupta1}\cite{Skadhauge1}\cite{Beacom1} by
detecting SN neutrinos on the Earth. Several neutrino experiments,
including the Daya Bay reactor neutrino underground laboratory
\cite{Guo2}, can be used to detect neutrino events from a SN
explosion and serve as a SN Early Warning System \cite{Antonioli}.

Recently, several methods to obtain information on neutrino parameters
by the study of SN neutrinos have been proposed. In
Ref. \cite{Dasgupta1}, Dasgupta {\it et al.} proposed a method to
identify the neutrino mass hierarchy for very small $\theta_{13}$
($\sin^2\theta_{13}\leq10^{-5}$) through the Earth matter effects.
In Ref. \cite{Skadhauge1}, Skadhauge {\it et al.} proposed methods
to identify the neutrino mass hierarchy for larger $\theta_{13}$
($\sin^2\theta_{13}\geq10^{-4}$) and obtain information about
$\theta_{13}$ by SN neutrinos. In Ref. \cite{Beacom1},
using the event number of the delayed SN neutrinos,
a method to measure neutrino masses was proposed by Beacom {\it et
al.}. In these works, the SN shock effects and the collective
effects were not taken into account. Furthermore, uncertainties
of some of the parameters appearing in SN neutrino fluxes, such as the
luminosities $L^0_{\alpha}$, temperatures $T_{\alpha}$, and pinching
parameters $\eta_{\alpha}$, were not considered.

In Ref. \cite{Guo3}, we studied the Earth matter effects in the
detection of SN neutrinos at the Daya Bay experiment and proposed a
possible method to acquire information about the neutrino mixing
angle $\theta_{13}$ smaller than $1.5^{\circ}$ which is difficult to
access in reactor neutrino experiments. We defined the ratio $R$ as
the event number of $\nu_e$ over that of $\bar{\nu}_e$. Using the
relation between the event number of SN neutrinos $N$ and
$\theta_{13}$, we can obtain information about small $\theta_{13}$
by measuring $R$. In that paper, a simplified picture for the
collective effects was used \cite{Dasgupta2} and the neutrino flavor
transformation due to the SN shock effects was not considered.
Furthermore, it was assumed that different flavor SN neutrinos have
the same time-independent luminosity. Hence in that calculation
there are several uncertainties which we will address in the present
work.

In this paper, we study the collective effects more accurately
\cite{Kneller1}\cite{Duan1} and consider the neutrino flavor
transformation arising from the SN shock effects
\cite{Wilson1}\cite{Schirato1}\cite{Fogli1}. In addition, according
to current Monte Carlo simulations on SN neutrino spectra
\cite{Keil1}, we allow different flavor neutrinos to have different
time-dependent luminosities. Furthermore, in order to reduce the
uncertainties of unknown parameters in SN neutrino fluxes such as
luminosity, temperature, and pinching parameter, we will define the
ratio of the event number of SN neutrinos for the charged-current
reaction detected after one second over that detected before one
second (this is different from the ratio considered in Ref.
\cite{Guo3}). With this ratio, we may identify the mass hierarchy,
$\Delta m^2_{31}>0$ vs $\Delta m^2_{31}<0$, and acquire information
about $\theta_{13}$. This method can be applied to neutrino
detectors of various types. Examples are the liquid scintillator
detector of the Daya Bay experiment \cite{Guo2} under construction,
the decommissioned heavy water detector of the previous Sudbury
Neutrino Observatory (SNO) \cite{SNO1}, etc.

The other main purpose of the present work is to propose a method to
obtain some information about neutrino masses from SN neutrinos.
Since neutrinos have small masses, their transmit velocities are
smaller than that of light. Therefore, a massive neutrino will have
an energy and mass dependent delay relative to a massless neutrino
after traveling over a long distance \cite{Beacom1}. We will define
another ratio, for the neutral-current reactions, as the event
number of the delayed SN neutrinos over the total event number. With
the relation between this ratio and neutrino masses, we will propose
a possible method to acquire information about neutrino masses. This
method will also be applied to Daya Bay and SNO for examples.

The paper is organized as follows. In Sec. II, we review the
detection of SN neutrinos on the Earth. In Sec. III, we propose a
possible method to identify the mass hierarchy and acquire
information about $\theta_{13}$ and apply the method to the Daya Bay
and SNO experiments. In Sec. IV, a method to acquire information
about neutrino masses is given and also applied to Daya Bay and SNO.
A summary is given in section V.

\section{\label{sec:hamckm}detection of SN neutrinos on the Earth}

In the explosion a typical type II SN generates intensive neutrinos
which carry about $99\%$ of the total energy and the explosion
itself consumes about 1$\%$ of the total energy. The vast amount of
neutrinos are produced in two bursts. In the first burst which lasts
for only a few milliseconds, electron neutrinos are generated by the
electron capture by nuclei $e^-+N(Z,A)\rightarrow N(Z-1,A)+\nu_e$
via the inverse beta decay $e^-+p\rightarrow n+\nu_e$ which leads to
a neutron rich star. In the second burst which lasts longer (about
10 seconds), neutrinos of all flavors are produced through the
electron-positron pair annihilation $e^-+e^+\rightarrow
\nu_{\alpha}+\bar{\nu}_{\alpha}$, the electron-nucleon
bremsstrahlung $e^{\pm}+ N\rightarrow
e^{\pm}+N+\nu_{\alpha}+\bar{\nu}_{\alpha}$, the nucleon-nucleon
bremsstrahlung $N+N\rightarrow N+N+\nu_{\alpha}+\bar{\nu}_{\alpha}$,
the plasmon decay $\gamma\rightarrow
\nu_{\alpha}+\bar{\nu}_{\alpha}$, and the photoannihilation
$\gamma+e^{\pm}\rightarrow e^{\pm}+\nu_{\alpha}+\bar{\nu}_{\alpha}$
\cite{book}.

We assume a "standard" SN explosion at a distance $D=10kpc$ from the
Earth, releasing a total energy $E_B=3\times10^{53}erg$ (similar to
SN1987A) and the
luminosity flux of the SN neutrino 
$\alpha$, $L_{\alpha}$, distributed in time as
\cite{Spergel1}\cite{Loredo1}\cite{Totsuka}
\begin{equation}
L_{\alpha}(t)=L^0_{\alpha}e^{-t/\tau}, \label{L}
\end{equation}
where $L^0_{\alpha}$ is the luminosity of the neutrino $\alpha$ at
$t=0$ and $\tau$ the luminosity decay time. The following range of
$\tau$ was obtained by fitting the the experimental data of SN1987A
with the exponentially decaying luminosity
\cite{Spergel1}\cite{Loredo1}:
\begin{equation}
\tau=1.74-4.19 s. \label{tau}
\end{equation}
The typical relations among $L^0_{\alpha}$ obtained from numerical
simulations are \cite{Keil1}
\begin{equation}
 \frac{L^0_{\nu_{e}}}{L^0_{\nu_{x}}}=(0.5-2), \quad L^0_{\nu_{e}}=L^0_{\bar{\nu}_{e}},
 \quad L^0_{\nu_{x}}=L^0_{\bar{\nu}_{x}},\label{L0}
  \quad  (x=\mu, \tau).
\end{equation}

There are two parametrizations of SN neutrino fluxes used in general
simulations, one used by the Livermore group
\cite{Wilson2}\cite{Totani1} and the other by the Garching group
\cite{Keil1}. In the present work, we use the Livermore
parameterization which is similar to the Fermi-Dirac distribution.
For the SN neutrino of flavor $\alpha$, the time-integrated neutrino
energy spectra is given by \cite{Totani1}
\begin{equation}
 F_{\alpha}^{(0)}(E)=\frac{L_{\alpha}}{F_{\alpha
 3}T_{\alpha}^{4}}\frac{E^2}{\exp{(E/T_{\alpha}-\eta_\alpha)}+1},
 \label{Foa}
\end{equation}
where $E$ is the neutrino energy, $T_{\alpha}$ the temperature of
the neutrino $\alpha$, $\eta_\alpha$ the pinching parameter of the
spectrum, $L_{\alpha}$ the luminosity given by Eq. (\ref{L}), and
$F_{\alpha j}$ is defined by
\begin{equation}
 F_{\alpha j}=\int_{0}^{\infty}\frac{x^j}{\exp{(x-\eta_\alpha)}+1}{\rm
 d}x, \nonumber
\end{equation}
where $j$ is an integer. The spectra obtained from numerical
simulations can be well fitted by \cite{Janka1}
\begin{eqnarray}
 T_{\nu_e}=3-4MeV, & T_{\bar{\nu}_e}=5-6MeV, & T_{\nu_x}=T_{\bar{\nu}_x}=7-9MeV,
 \label{T}\\
 \eta_{\nu_e}\approx3-5, & \eta_{\bar{\nu}_e}\approx2.0-2.5, &
 \eta_{\nu_x}=\eta_{\bar{\nu}_x}\approx0-2. \nonumber
 \end{eqnarray}

While neutrinos propagate outward to the surface of the SN they
could be subjected to the SN shock effects
\cite{Schirato1}\cite{Fogli1}\cite{Fogli2}\cite{Lunardini1}\cite{Takahashi1},
the MSW effects \cite{MSW1}\cite{Kuo1}\cite{Smirnov1}, and the
collective effects
\cite{Dasgupta2}\cite{Duan1}\cite{Duan2}\cite{Duan3}\cite{Duan4}\cite{Duan5}\cite{Hannestad1}\cite{Raffelt1}\cite{Fogli3}.
Then, after travelling the cosmic distance to reach the Earth, they
go through a part of the Earth and are subjected to the Earth matter
effects \cite{Dighe2}\cite{Guo4}\cite{Ioannisian1}\cite{Lunardini2}.
In the following, we will consider the above four types of effects
in detail.

\subsection{\label{subsec:form}Time dependence of the signal: SN shock effects }

As was pointed out in Ref. \cite{Schirato1}, the shock propagating
inside the SN may reach the region of densities relevant for
neutrino conversion during the period of neutrino emission. It
modifies the density profile of the star, thus affecting neutrino
flavor transitions after the core bounce
\cite{Schirato1}\cite{Fogli1}. The time-dependent variations of the
neutrino potential in matter along with the SN shock profile can
leave distinctive imprints on the energy and time structure of the
neutrino signal. The study of such effects may provide us with
additional constraints for the determination of the neutrino mass
hierarchy, the neutrino mixing angle $\theta_{13}$, and neutrino
masses.

In Ref. \cite{Fogli1}, the authors introduce a simplified, empirical
parametrization of the shock density profile $\rho$, which
reproduces the main features of the graphical profile of Ref.
\cite{Schirato1} and is continuous in both $t$ and the supernova
radial coordinate $r$. In this way, numerical and analytical
calculations of the crossing probability at the high resonance
region $P_H$ can continuously cover the relevant parameter space. We
follow the notation of Ref. \cite{Fogli1} in the discussion below.

For post-bounce time $t<1s$ \cite{Bethe1} before the shock effects
begin, the SN matter density is
\begin{equation}
 \rho_0(r)\backsimeq 10^{14}\cdot\Bigg(\frac{r}{1km}\Bigg)^{-2.4} g/cm^3.
 \label{rho0}
\end{equation}
For $t\geqslant1s$ the shock effects set in and the matter density
is given by
\begin{equation}
\rho(r,t)=\rho_0(r)\cdot
\begin{cases}\xi\cdot f(r,t) & (r\leqslant r_s), \\
 1 & (r>r_s),
 \end{cases}
\label{rho}
\end{equation}
where $f(r,t)$ is defined as
\begin{equation}
 f(r,t)=\exp\{[0.28-0.69\ln(r_s/km)][\arcsin(1-r/r_s)]^{1.1}\},
 \label{f}
\end{equation}
and $\xi$ is a typical ratio of the potential across the shock
front,
\begin{eqnarray}
 \xi=V_+/V_-\backsimeq10, \label{xi}
\end{eqnarray}
which measures the SN matter potential $V(r)$ drop from
\begin{eqnarray}
V_+=\lim_{r\rightarrow r_s^-}V(r) \label{V+}
\end{eqnarray}
to
\begin{eqnarray}
V_-=\lim_{r\rightarrow r_s^+}V(r). \label{V-}
\end{eqnarray}
The SN matter potential is related to the SN electron density
$N_e(r)$ by
\begin{eqnarray}
V(r)=\sqrt{2}G_F N_e(r)=\sqrt{2}G_F N_A\rho(r) Y_e \label{VS},
\end{eqnarray}
where $N_A$ is the Avogadro's number and $Y_e$ is the electron
fraction. In the numerical calculations, we assume $Y_e=0.5$. In
Eqs. (\ref{rho0})-(\ref{xi}), a slightly accelerating shock-front
position $r_s$ is assumed with the explicit time dependence,
\begin{eqnarray}
 r_s(t)=-4.6\times10^{3}+1.13\times10^{4}\cdot t+1\times10^{2}\cdot t^2,
  \label{rs}
\end{eqnarray}
where $r_s$ is in units of $km$ and $t$ in units of $s$.

Using Eqs. (\ref{rho0})-(\ref{xi}), we plot in Fig. 1 the density
profile of the SN as a function of the radius at different times.
These curves reproduce reasonably well the main features of the
shock profiles as shown in Ref. \cite{Schirato1}. The straight line
labelled by $t=0$ is the unmodified density profile given by Eq.
(\ref{rho0}). Once the shock effects begin, at a given time the
shock radius $r_s$ is determined by Eq. (\ref{rs}). The density
profile is suppressed more and more in the region $r<r_s$ as the
time and therefore $r_s$ increase. The density reaches  a minimum
and then rises relatively quickly to the value $\xi\rho_0(r_s)$ as
$r\rightarrow r_s^-$. In the region $r>r_s$ the density profile is
the unperturbed original form $\rho_0$. The density suffers a
sizable discontinuity at $r_s$.\footnote{It can be shown that
$f(r_s,t)=1$ and therefore the discontinuity is $(\xi
-1)\rho_0(r_s)\approx 9\rho(r_s)$.} For neutrinos emitted in the
region $r>r_s$, due to their much higher velocities, which is the
order of speed of the light, than that of the shock wave
propagation, which is the order of $10^4~km/s$, they are not
influenced by the shock effects. Hence the majority of SN neutrinos
which are emitted within a 10 second period will be able to semble
the whole range of the profile of the SN density under the influence
of the shock. An important consequence of the density modification
by the shock wave is that some of the neutrinos can undergo multiple
resonance flavor conversion when they traverse the SN. We consider
them below.

Let us first consider the neutrino flavor conversion due to the
density discontinuity at the shock front $r_s$.  This crossing
probability of the neutrino flavor transformation which we denote as
$P_s$ can be determined by the conservation of flavor across the
matter potential discontinuity. By a simple calculation, we have
\cite{Kuo1}
\begin{equation}
 P_s=\sin^2(\theta^+_m-\theta^-_m),
 \label{Ps}
\end{equation}
where $\theta^+_m$ and $\theta^-_m$ are respectively the effective
mixing angles in matter immediately before and after the shock front
position $r_s$. Such angles are defined by
\begin{equation}
\cos2\theta^{\pm}_m=\frac{\Delta m^2_{31} \cos2\theta_{13}-
 2EV^{\pm}}{\sqrt{(\Delta m^2_{31} \cos2\theta_{13}-2EV^{\pm})^2
 +(\Delta m^2_{31} \sin2\theta_{13})^2}}.
 \label{thetam}
\end{equation}

\subsection{\label{subsec:form}MSW effects}

The MSW effects \cite{MSW1} is the adiabatic or partially adiabatic
neutrino flavor conversion in media with locality or time varying
matter density. The main feature is that in a high density medium a
flavor is also a mass state which propagates without flavor
oscillation, until crossing a resonance where there is a finite
probability for flavors change, i.e., a jump from one mass
eigenstate to another. For $\theta_{13}$ smaller than $3^o$, using
the Landau-Zener formula \cite{Landau1}, one can obtain the crossing
probability, $P_{Hi}$, for neutrinos to jump from one mass
eigenstate to another at the high resonance layer ($r=r_i$)
\begin{eqnarray}
P_{Hi}=\exp(-\frac{\pi}{2}\gamma), \label{PH0}
\end{eqnarray}
with
\begin{eqnarray}
\gamma=\frac{|\Delta m_{31}^2|
}{2E}\frac{\sin^22\theta_{13}}{\cos2\theta_{13}}\frac{1}{|d\ln
N_e/dr|_{res}}, \label{gamma}
\end{eqnarray}
where $N_e$ is the electron density \cite{Kuo1} and $|\Delta
m^2_{31}|\backsimeq|\Delta m^2_{32}|\backsimeq2.4\times10^{-3} eV^2$
\cite{Gonzalez-Garcia1}. Similarly, one can calculate the expression
of the crossing probability $P_L$ at the low resonance region inside
the SN. However, due to the large angle solution of the solar
neutrino, $P_L\simeq0$.

In the presence of a propagating shock wave, the calculation of the
global crossing probability, $P_H$, is not straightforward. The
shock modifies the density profile as discussed in the preceding
subsection. Within the shock wave and near the shock front the
matter density is no longer monotonically decreasing, but rather
goes through a minimum and a maximum. So the resonance condition can
be satisfied possibly as many as three points:
\begin{eqnarray}
\Delta m^2_{31}\cdot cos2\theta_{13}=2\sqrt{2}G_FN_e(r_i)E,\quad
(i=1,2,3). \label{resonance}
\end{eqnarray}
It is, of course, possible that there are only two or one resonance
points, or even none at all. In the case of three resonance points,
which we denote as $r_i,~ i=1,~2,~3$, two of the points lie below
the shock front and the third one beyond the shock front, i,e.,
$r_1<r_2<r_s<r_3$. The crossing probability $P_{Hi}$ at $r_i$,
$i=1,~2,~3$, can be calculated from Eq. (\ref{PH0}). Note that
although the matter densities are the same at all resonance points
as required by Eq. (\ref{resonance}), their crossing probabilities,
as given by Eq. (\ref{gamma}), may be different because the slop of
the matter density may be different at different resonance points.
The global crossing probability can be expressed in terms of four
crossing probabilities $P_{H1}$, $P_{H2}$, $P_{s}$, and $P_{H3}$.

To obtain the expressions of the global crossing probabilities, we
first define four densities: $\rho_{res}$ is the resonance density
which satisfies Eq. ({\ref{resonance}}), $\rho_{+}$ the density at
$r=r_s$ corresponding to the matter potential $V_+$, $\rho_{-}$ the
density at $r=r_s$ corresponding to the matter potential $V_-$, and
$\rho_b$ the density at the bottom of the camber which is the
minimum of the density profile below the shock front. From Fig. 1,
it can be seen that
\begin{eqnarray}
 \rho_+>\rho_-, \quad \rho_+>\rho_b. \nonumber
\end{eqnarray}
In the pre-bounce time period $t<1s$, the resonance condition occurs
at only one point and there are no neutrino shock effects. Then,
\begin{eqnarray}
P_H=P_{H3}.\label{PH1}
\end{eqnarray}
In the post-shock period $t\geq1s$ we have four different
situations:

(1) For $\rho_-<\rho_{res}<\rho_b$, the resonance condition doesn't
occur and then
\begin{eqnarray}
P_H=P_s.   \label{PH2}
\end{eqnarray}

(2) For $\rho_{res}>\rho_+$, or $\rho_{res}=\rho_b$ and
$\rho_{res}>\rho_-$, or $\rho_{res}<\rho_b$ and
$\rho_{res}\leq\rho_-$, the resonance condition occurs at only one
point, $r=r_1$ or $r=r_3$, and then
\begin{eqnarray}
P_H=P_{Hk}+P_{s}-2P_{Hk}P_{s}+2\sqrt{P_{Hk}P_{s}(1-P_{Hk})(1-P_{s})}
\cos\phi_{ks}, \quad (k=1, 3). \label{PH3}
\end{eqnarray}

(3) For $\rho_b<\rho_{res}\leq\rho_+$ and $\rho_{res}>\rho_-$, or
$\rho_{res}=\rho_b$ and $\rho_{res}\leq\rho_-$, the resonance
condition occur at two points, $r=r_1, r_2$ or $r=r_1, r_3$, and
then
\begin{eqnarray}
P_H&=&(P_{H1}+P_{Hl}+P_{s})-2(P_{H1}P_{Hl}+P_{H1}P_{s}+P_{Hl}P_{s})+4P_{H1}P_{Hl}P_{s}
\nonumber\\
&&+2(1-2P_s)\sqrt{P_{H1}P_{Hl}(1-P_{H1})(1-P_{Hl})}\cos\phi_{1l}
\label{PH4}\\
&&+2(1-2P_{H1})\sqrt{P_{Hl}P_{s}(1-P_{Hl})(1-P_{s})}\cos\phi_{ls}
\nonumber\\
&&+2(1-2P_{Hl})\sqrt{P_{H1}P_{s}(1-P_{H1})(1-P_{s})}\cos\phi_{1s}
\quad (l=2, 3).   \nonumber
\end{eqnarray}

(4) For $\rho_b<\rho_{res}\leq\rho_-$, the resonance condition occur
at three points, $r=r_1, r_2, r_3$, and then
\begin{eqnarray}
P_H&=&(P_{H1}+P_{H2}+P_{s}+P_{H3})-2(P_{H1}P_{H2}+P_{H1}P_{H3}+P_{H1}P_{s}
+P_{H2}P_{H3}+P_{H2}P_{s}+P_{H3}P_{s}) \nonumber\\
&&
+4(P_{H1}P_{H2}P_{H3}+P_{H1}P_{H2}P_{s}+P_{H1}P_{H3}P_{s}+P_{H2}P_{H3}P_{s})-8P_{H1}P_{H2}P_{H3}P_s
\nonumber\\
&&+2(1-2P_s-2P_{H3}+4P_{H3}P_s)\sqrt{P_{H1}P_{H2}(1-P_{H1})(1-P_{H2})}\cos\phi_{12}
\nonumber\\
&&+2(1-2P_{H1}-2P_{H3}+4P_{H1}P_{H3})\sqrt{P_{H2}P_{s}(1-P_{H2})(1-P_{s})}\cos\phi_{2s}
\nonumber\\
&&+2(1-2P_{H2}-2P_{H3}+4P_{H2}P_{H3})\sqrt{P_{H1}P_{s}(1-P_{H1})(1-P_{s})}\cos\phi_{1s}
\label{PH5}\\
&&+2(1-2P_{H1}-2P_{H2}+4P_{H1}P_{H2})\sqrt{P_{H3}P_{s}(1-P_{H3})(1-P_{s})}\cos\phi_{s3}
\nonumber\\
&&+2(1-2P_s-2P_{H1}+4P_{H1}P_s)\sqrt{P_{H2}P_{H3}(1-P_{H2})(1-P_{H3})}\cos\phi_{23}
\nonumber\\
&&+2(1-2P_s-2P_{H2}+4P_{H2}P_s)\sqrt{P_{H1}P_{H3}(1-P_{H1})(1-P_{H3})}\cos\phi_{13}
\nonumber\\
&&-8\sqrt{P_{H1}P_{H2}(1-P_{H1})(1-P_{H2})}\cos\phi_{12}\sqrt{P_{H3}
P_{s}(1-P_{H3})(1-P_{s})}\cos\phi_{s3}.  \nonumber
\end{eqnarray}
In Eqs. (\ref{PH3}-\ref{PH5}), $\phi_{ij}$ is defined as
\begin{eqnarray}
\phi_{ij}\approx\int^{r_j}_{r_i}dx\frac{1}{2E}\sqrt{[\Delta{m}^2_{31}
   \cos2\theta_{13}-2EV(r)]^2+(\Delta{m}^2_{31} \sin2\theta_{13})^2},
\label{phiij}
\end{eqnarray}
with $V(r)$ being the SN matter potential. A summary for neutrino
flavor conversions due to the neutrino shock effects and the MSW
effects in various density regions is given in Table I. Due to the
SN matter effects and the SN shock effects, the expressions of
$P_H$, Eqs. (\ref{PH1})-(\ref{PH5}), are different from Eq. (25) in
Ref. \cite{Fogli1} and Eq. (34) in Ref. \cite{Duan2}.

Because of the various effects they have been subjected to, the SN
neutrinos reaching the Earth have a very rich structures to be
explored in their spectra in time, energy, and the mixing angle
$\theta_{13}$.  In Fig. 2, we plot the crossing probability $P_H$ as
a function of the neutrino energy $E$ at $t=4s$ for two very small
mixing angles $\sin^2\theta_{13}=10^{-5}$ and $10^{-4}$. It is seen
that there are large variations in $P_H$ as the neutrino energy
varies. These forms of behavior are different from those given in
Refs. \cite{Guo3} and \cite{Fogli1}. In Fig. 3, we plot $P_H$ as a
function of $t$ for $\sin^2\theta_{13}=10^{-4}$ at a fixed $E$ for
three different values of $E$. It can be seen that the shapes of the
curves depend strongly on $E$ and there are many oscillations in
$P_H$ in an interval around $t=5s$.  The interval enlarges and moves
to higher $t$ value as $E$ increases. This is different from that
given in Ref. \cite{Fogli1}. In Fig. 4, we plot $P_H$ as a function
of $\theta_{13}$ for different values of $E$ and $t$. It is found
that at some values of $\theta_{13}$, which depends on the energy of
SN neutrinos, $P_H$ begins to change very rapidly with $\theta_{13}$
at $t=4s$, and the behavior of the $P_H$ dependence on $\theta_{13}$
is different from that given in Ref. \cite{Guo3}. These differences
are attributable to the different SN matter effects when the shock
effects is included in the consideration.  The relation between
$P_H$ and $\theta_{13}$ will provide a way to obtain information on
$\theta_{13}$ and the study of the shock wave effects.

\subsection{\label{subsec:form}Neutrino-neutrino interactions:
         collective effects}

Because of the neutrino-neutrino forward scattering or neutrino
self-interaction, neutrinos can experience collective flavor
transformation in the SN near the neutrinosphere where the neutrino
number densities are very large \cite{Duan4}. This phenomenon is
different from the MSW effects in that the flavor evolution
histories of neutrinos in the collective oscillations are coupled
together and must be solved simultaneously.

Recently, a significant amount of studies on collective effects
have been made by a number of authors, e.g.,
Duan {\it et al.} \cite{Duan2}\cite{Duan3}\cite{Duan4}\cite{Duan5},
Dasgupta {\it et al.} \cite{Dasgupta1}\cite{Dasgupta2}, Hannestad
{\it et al.} \cite{Hannestad1}, Raffelt {\it et al.}
\cite{Raffelt1}, and Fogli {\it et al.} \cite{Fogli3}. Defining
$P_{\nu\nu}$($\bar{P}_{\nu\nu}$) as the probability that the
(anti-)neutrino $\nu$($\bar{\nu}$) remains as $\nu$($\bar{\nu}$)
after the collective effects,  Dasgupta
{\it et al.} \cite{Dasgupta1} introduced a much simplified picture to explain the
characteristics of the collective effects which we followed in our
earlier paper \cite{Guo3}, i.e. $P_{\nu\nu}=\bar{P}_{\nu\nu}=1$ in
the case of normal mass hierarchy; and $\bar{P}_{\nu\nu}=1$,
\begin{equation}
P_{\nu\nu}=
\begin{cases} 1 & (E<E_C), \\
 0 & (E>E_C),
 \end{cases}
\nonumber
\end{equation}
in the case of inverted mass hierarchy, where $E_C$ is a critical
energy across which a stepwise flavor conversion of $\nu_e$
develops. However, from the numerical simulations of the collective
effects obtained by Duan {\it et al.} \cite{Duan1}, it is found that
$P_{\nu\nu}$($\bar{P}_{\nu\nu}$) depends on the neutrino
(anti-neutrino) energy $E$, the SN radial coordinate $r$, the SN
neutrino emission angle, and it is very difficult to obtain an
analytical expression for $P_{\nu\nu}$($\bar{P}_{\nu\nu}$). It can
be seen from Movies 5 and 6 in Ref. \cite{Duan1}\footnote{Movies 5
and 6 which describe the full simulation of the collectives effects
\cite{Duan1} are available at
http://iopscience.iop.org/1749-4699/1/1/015007/media.} that when the
SN radial coordinate $r$ increases, there are oscillations between
$\nu_e$ ( $\bar{\nu}_e$) and $\nu_{\tau}$ ($\bar{\nu}_{\tau}$) due
to the collective effects. When $r\geq220km$, which is far away from
the neutrinosphere, the collective effects for the neutrino amount to
a swap between the $\nu_e$ and $\nu_{\tau}$ fluxes, irrespective of
the mass hierarchy being normal or inverted.  For the anti-neutrino,
in the case of normal
mass hierarchy, the fluxes of $\bar{\nu}_e$ and $\bar{\nu}_{\tau}$
are nearly unchanged and there is no swap between them. In the case
of inverted mass hierarchy, the behaviors of the $\bar{\nu}_e$ and
$\bar{\nu}_{\tau}$ fluxes are very complex and, for simplicity, they
are generally approximated \cite{Kneller1}\cite{Duan4} as that there
is no swap between them. In this way, we apply the following new
simplified picture to describe the characteristics of the collective
effects:
\begin{equation}
P_{\nu\nu}=
\begin{cases} 1 & (E<E_C), \\
 0 & (E>E_C),
 \end{cases}
\label{Pnunu1}
\end{equation}
for neutrinos and
\begin{equation}
\bar{P}_{\nu\nu}=1, \label{Pnunu2}
\end{equation}
for antineutrinos. It will be seen from our numerical calculations
that the results obtained in Secs. III and IV hardly change for
$E_C$ varying between $6MeV$ and $10MeV$. Therefore, similar to Ref.
\cite{Kneller1}, we take $E_C=10MeV$.

\subsection{\label{subsec:form}Earth matter effects}

Neutrinos from a SN may travel through a significant portion of the
Earth before reaching the detector and are subjected to the matter
effects \cite{Dighe2}. Therefore, we need to consider the Earth
matter effects in studying SN neutrinos. Suppose a neutrino reaches
the detector at the incident angle $\theta$ (see Fig. 5). Then the
distance that the neutrino travels through the Earth to the detector
$L$ and the distance of the neutrino to the center of the Earth
$\tilde{x}$ are given by
\begin{eqnarray}
 L&=&(-R_E+h)\cos{\theta}+\sqrt{R_E^2-(R_E-h)^2\sin^2{\theta}},
 \nonumber\\
 \tilde{x}&=&\sqrt{(-R_E+h)^2+(L-x)^2+2(R_E-h)(L-x)\cos{\theta}},
 \nonumber
\end{eqnarray}
where $h$ is the depth of the detector in the Earth, $x$ the
distance that the neutrino travels into the Earth, and $R_E$ the
radius of the Earth.

Let $P_{ie}$ be the probability that a neutrino mass eigenstate
$\nu_i$ enters the surface of the Earth and arrives at the detector
as an electron neutrino $\nu_e$, we have \cite{Ioannisian1}
\begin{equation}
P_{2e}=\sin^2\theta_{12}+\frac{1}{2}\sin^22{\theta_{12}}\int_{x_0}^{x_f}dxV(x)\sin\phi
_{x\rightarrow x_f}^{m}, \label{P2e}
\end{equation}
where $V(x)$ is the potential $\nu_e$ experiences due to the matter
density $\rho(x)$ inside the Earth
\cite{Guo3}\cite{Ioannisian1}\cite{Earth}
\begin{eqnarray}
V(x)=\sqrt{2}G_F N_A \rho(x) Y_e, \label{VE}
\end{eqnarray}
and $\phi_{a\rightarrow b}^m$ is defined as
\begin{eqnarray}
\phi_{a\rightarrow b}^m &=& \int_a^b {\rm d}x \Delta_m(x),\nonumber
\end{eqnarray}
where
\begin{eqnarray}
\Delta_m(x)&=&\frac{\Delta m_{21}^2}{2E}
 \sqrt{(\cos2\theta_{12}-\varepsilon(x))^2+ \sin^22\theta_{12}},
 \label{detlam}
\end{eqnarray}
with $\theta_{12}\backsimeq34.5^{\circ}$, $\Delta
m^2_{21}=7.7\times10^{-5}eV^2$, $\varepsilon(x)={2EV(x)}/{\Delta
m_{21}^2}$ \cite{Gonzalez-Garcia1}.

\subsection{\label{subsec:form}Summary of all matter effects}

When all the effects, including the shock, MSW, collective, and the
Earth matter, are taken into account, the SN neutrino fluxes at the
detector can be written as
\begin{eqnarray}
F_{\nu_e}^D&=&pF_{\nu_e}^{(0)}+(1-p)F_{\nu_x}^{(0)},
\nonumber \\
F_{\bar{\nu}_e}^D&=&\bar{p}F_{\bar{\nu}_e}^{(0)}+
  (1-\bar{p})F_{\bar{\nu}_x}^{(0)},
\nonumber \\
2F_{\nu_x}^D&=&(1-p)F_{\nu_e}^{(0)}+(1+p)F_{\nu_x}^{(0)},
\nonumber \\
2F_{\bar{\nu}_x}^D &=&(1-\bar{p})F_{\bar{\nu}_e}^{(0)}+
 (1+\bar{p})F_{\bar{\nu}_x}^{(0)},
\label{FD}
\end{eqnarray}
where the survival probabilities $p$ and $\bar{p}$ are given by
\begin{eqnarray}
p&=&P_{2e}[P_HP_{\nu\nu}+(1-P_H)(1-P_{\nu\nu})], \nonumber\\
\bar{p}&=&(1-\bar{P}_{2e})\bar{P}_{\nu\nu}, \label{pn}
\end{eqnarray}
for the normal mass hierarchy and
\begin{eqnarray}
p&=&P_{2e}P_{\nu\nu}, \nonumber\\
\bar{p}&=&(1-\bar{P}_{2e})[\bar{P}_H\bar{P}_{\nu\nu}+
  (1-\bar{P}_H)(1-\bar{P}_{\nu\nu})],
\label{pi}
\end{eqnarray}
for the inverted mass hierarchy.

Therefore, the event numbers $N(i)$ of SN neutrinos in the various
reaction channels "$i$" can be calculated by integrating over the
neutrino energy, the product of the target number $N_T$, the cross
section of the given channel $\sigma(i)$, and the neutrino flux
function at the detector $F_{\alpha}^D$,
\begin{equation}
 N(i)=N_T\int{{\rm d}E\cdot\sigma(i)\cdot\frac{1}{4\pi
 D^2}\cdot F_{\alpha}^D}, \label{Ntotal}
\end{equation}
where $\alpha$ stands for the neutrino or antineutrino of a given
flavor, and $D$ is the distance between the SN and the Earth. In the
following section,
we propose a method to identify the mass hierarchy and acquire
information about $\theta_{13}$ by detecting SN neutrinos.

\section{\label{sec:form}Identify the mass hierarchy and acquire
         information about $\theta_{13}$ by detecting SN neutrinos}

In the above section, we have discussed the SN shock effects, the
MSW effects, the collective effects, and the Earth matter effects in
detail for the detection of SN neutrinos. From Eqs.
(\ref{Ps})-(\ref{gamma}) and (\ref{PH1})-(\ref{Ntotal}), it can be
seen that the event number of SN neutrinos depends on the mixing
angle $\theta_{13}$ for different mass hierarchies. Therefore, we
may identify the mass hierarchy and acquire information about
$\theta_{13}$  by detecting SN neutrinos. In Ref. \cite{Guo3}, we
defined a ratio of the event number of $\nu_e$ over that of
$\bar{\nu}_e$ and studied the method to acquire information about
$\theta_{13}$. However, in that paper, we ignored the SN shock
effects and only a very simple model for the collective effects was
considered. In addition, with the naive luminosity of SN neutrinos
employed there, the uncertainties in $\theta_{13}$ will be very
large. In the present work, we take into account the SN shock
effects and consider the collective effects more realistically. In
order to further reduce the influence of the uncertainties in the
parameters of the SN neutrino fluxes, we define a more appropriate
ratio as the event number of SN neutrinos for the charged-current
reaction detected after $1s$ over that detected before $1s$. Using
this new ratio, we can better acquire information about
$\theta_{13}$ and identify the neutrino mass hierarchy. From Ref.
\cite{Guo3} and detailed numerical studies related to the present
work, we found that the ratios of event numbers of SN neutrinos are
nearly independent of the SN neutrino incident angle. Therefore, in
the following we will only consider the case for the incident angle
of $30^\circ$ .

There will be eight detectors located at the near and far sites of
the Daya Bay experiment \cite{Guo2}. The total detector mass is
about 300 tons and the depth of the detector $h\backsimeq400$ m. The
Daya Bay Collaboration has decided to use Linear Alkyl Benzene (LAB)
as the main part of the liquid scintillator. LAB has a chemical
composition including $C$ and $H$ and the ratio of the number of $C$
and $H$ is about 0.6. Then, the total numbers of target protons,
electrons, and $^{12}C$ are
\begin{eqnarray}
N_T^{(p)}=2.20\times10^{31}, \quad N_T^{(e)}=1.01\times10^{32},
\quad N_T^{(C)}=1.32\times10^{31} \nonumber.
\end{eqnarray}

In the Daya Bay experiment, there are three reactions which can be
used to detect SN neutrinos: the inverse beta decay,
neutrino-electron reactions, and neutrino-carbon reactions. However,
since the event number observed in the channel of neutrino-electron
reactions at Daya Bay is very small \cite{Guo3}, these reactions
will not be considered in the present work. Therefore, we only
consider the inverse beta decay and the neutrino-carbon reactions,
their effective cross sections are given as follows
\cite{Cadonati1}\cite{Burrows1}:

(1) The inverse beta decay is

\begin{eqnarray}
 \sigma(\bar{\nu}_ep)=9.5\times10^{-44}(E(MeV)-1.29)^2 cm^2, \label{sigma p}
\end{eqnarray}
where the reaction threshold is $E_{th}=1.80MeV$.

(2) The neutrino-carbon reactions are

\begin{eqnarray}
 &&\ \langle\sigma(^{12}C(\nu_e,e^{-})^{12}N)\rangle=1.85\times10^{-43}
cm^2,\nonumber \\
  &&\ \langle\sigma(^{12}C(\bar{\nu}_e,e^{+})^{12}B)\rangle=1.87\times10^{-42}
cm^2, \label{sigma CB}
\end{eqnarray}
for the charged-current capture,  and
\begin{eqnarray}
 &&\ \langle\sigma(\nu_e^{12}C)\rangle=1.33\times10^{-43}cm^2, \nonumber \\
 &&\ \langle\sigma(\Bar{\nu}_e^{12}C)\rangle=6.88\times10^{-43}cm^2,
 \label{sigma CC} \\
 &&\
 \langle\sigma(\nu_{\mu,\tau}(\Bar{\nu}_{\mu,\tau})^{12}C)\rangle=3.73\times10^{-42}cm^2,
 \nonumber
\end{eqnarray}
for the neutral-current capture. The effective cross sections in Eq.
(\ref{sigma CB}) are valid for SN neutrinos without oscillations.
When neutrino oscillations are taken into account, the oscillations
of higher energy $\nu_x$ into $\nu_e$ result in an increased event
rate since the expected $\nu_e$ energies in the absence of
oscillations are just at or below the charged-current reaction
threshold. This leads to an increase by a factor of 35 for the cross
section $\langle\sigma(^{12}C(\nu_e,e^-)^{12}N)\rangle$ if we
average it over a $\nu_e$ distribution with $T=8MeV$ rather than
$3.5MeV$. Similarly, the cross section
$\langle\sigma(^{12}C(\Bar{\nu}_e,e^+)^{12}B)\rangle$ is increased
by a factor of 5.

In order to reduce the uncertainties of the unknown parameters
appearing in the SN neutrino fluxes, including those in the
luminosities $L^0_{\alpha}$, temperatures $T_{\alpha}$, and pinching
parameters $\eta_{\alpha}$, we define the ratio $R^{+(-)}_1$ as the
event number of $\nu_e$($\bar{\nu}_e$) for the charged-current
reaction detected after $1s$ over that detected before $1s$,
\begin{eqnarray}
 R^{+(-)}_1=\frac{N^{CC}_{\nu_e(\bar{\nu}_e)}(1s<t<10s)}
   {N^{CC}_{\nu_e(\bar{\nu}_e)}(0<t<1s)}.\label{ratio1}
\end{eqnarray}
With Eqs. (\ref{L}), (\ref{Foa}), (\ref{Ps})-(\ref{gamma}),
(\ref{PH1})-(\ref{Ntotal}), and (\ref{ratio1}), we can see that
$R^{\pm}_1$ depends on the mixing angle $\theta_{13}$ and the
luminosity decay time $\tau$. Form a detailed numerical calculation,
we found that $R^{\pm}_1$ which changes with $\theta_{13}$ has also
a significant dependence on $\tau$.  This can lead to large
uncertainties in $R^{\pm}_1$ when $\tau$ varies in the range given
in Eq. (\ref{tau}). Therefore, it will be difficult to identify the
mass hierarchy and acquire information on $\theta_{13}$ from
$R^{\pm}_1$ when $\tau$ has a significant uncertainty. However, if
the value of $R^{\pm}_1$ is determined through SN neutrinos, we can
obtain a rather tight relation between $\tau$ and $\theta_{13}$. We
demonstrate this relation in Fig. 6, for the neutrino-carbon
reactions.  In Fig. 6 we plot $\tau$ as a function of $\theta_{13}$
for a fixed value of $R^+_1$.  We take $R^+_1=2.4$ and
$L^0_{\nu_e}/L^0_{\nu_x}=1$ as an example\footnote{We have checked
that when $R^{\pm}_1$ and $L^0_{\nu_e}/L^0_{\nu_x}$ take other
values, the qualitative results do not change.}. From this figure,
we see that the value of $\tau$ lies in a small range, i.e.,
$2.68-3.00$ when $\theta_{13}$ varies from 0 to $3^\circ$ for the
normal mass hierarchy.  This is much smaller than the presently
allowed range given in Eq. (\ref{tau}). It should be remarked that
several authors \cite{Kneller1}\cite{Schirato1}\cite{Fogli1} have
typically used the best fit value, i.e., $\tau=3s$ \cite{Loredo1} in
their study of SN neutrinos.

First consider the channel of the inverse beta decay in which only
$R^-_1$ is relevant. We plot $R^-_1$ in Fig. 7 as a function of
$\theta_{13}$ for the two mass hierarchies, and for
$L^0_{\nu_e}/L^0_{\nu_x}$, $T_{\alpha}$, and $\eta_{\alpha}$ taking
their allowed extreme values in the ranges given in Eqs. (\ref{L0})
and (\ref{T}).  For the normal mass hierarchy $R^-_1$, having the
value of 2.4, is independent of $\theta_{13}$. So it is impossible
to obtain any information on $\theta_{13}$ for the normal mass
hierarchy in the inverse beta decay process. For the inverted mass
hierarchy, for a given value of $L_{\bar{\nu}_e}/L_{\bar{\nu}_x}$,
$R^-_1$ lies in a band bounded by the curves marked as "inverted \&
max" and "inverted \& min", which are obtained by setting the values
of both $T_\alpha$ and $\eta_\alpha$ at their upper and lower
bounds, respectively. From this plot, it can be seen that if the
value of $R^-_1$ is away from 2.4, that is statically significant,
the mass hierarchy must be inverted and a band of values of
$\theta_{13}$ and $L_{\bar{\nu}_e}/L_{\bar{\nu}_x}$ can be
determined. But it is difficult to determine $\theta_{13}$ from
$R^-_1$ alone. It is interesting to observe that, since the value of
$R^-_1$ falls bound in the limited range of 2.1-2.8, it can serve as
a test of the validity of the overall parametrization of the SN
explosion mechanism.

For the neutrino-carbon reactions, both $R^+_1$ and $R^-_1$ can be
used to identify the mass hierarchy and obtain information on
$\theta_{13}$. In Fig. 8a, we plot $R^+_1$ vs $\theta_{13}$ similar
to Fig. 7. Here for the inverted mass hierarchy, $R^+_1$ has the
constant value 2.4 independent of $\theta_{13}$.  If $R^+_1$ is
larger than 2.4, the mass hierarchy must be normal. It can also be
seen from Fig. 8a that $R^+_1$ depends on $\theta_{13}$ sensitively
for the normal mass hierarchy and the uncertainties of $R^+_1$ due
to $L^0_{\alpha}$, $T_{\alpha}$, and $\eta_{\alpha}$ are small.
Therefore, in the case of normal mass hierarchy, we can restrict
$\theta_{13}$ to a small range of values by measuring $R^+_1$.
Similar plots are made for $R^-_1$ in Fig. 8b.  Now for the normal
mass hierarchy, $R^-_1$ is independent of $\theta_{13}$. For the
inverted mass hierarchy $R^-_1$ varies with $\theta_{13}$ and is
smaller than 2.4. Hence we may also be able to identify the mass
hierarchy by $R^-_1$. However, as shown in the plot, in the case of
inverted mass hierarchy the large uncertainties of $R^-_1$ due to
$L^0_{\alpha}$, $T_{\alpha}$, and $\eta_{\alpha}$ make it difficult
to constrain $\theta_{13}$ by $R^-_1$.

The Daya Bay experiment has the sensitivity limit of 0.01 for
$\sin^22\theta_{13}$, i.e., $3^\circ$ for $\theta_{13}$. However, if
a SN explosion takes place during the operation of the experiment,
it is possible to reach a much smaller value of $\theta_{13}$ with
the benefit of determining the mass hierarchy by measuring $R^+_1$
for the neutrino-carbon reactions.

Now let us consider the heavy water detector, for an example, used
in the previous SNO experiment \cite{SNO1}. The detector mass was
made of 1000 tons of heavy water located in the depth of $h=6000$
m.w.e.\footnote{m.w.e. means meter-water-equivalent and 1 m of rock
is about 2.7 m of water.} \cite{SNO1}\cite{Beacom2}.  We consider
only the neutrino-deuterium reactions. The total number of deuterium
is $N_T^{(d)}=6.02\times10^{31}$. Scaling the experimentally
measured energy values from the decay of the muon at rest to the
energy scale of SN neutrinos, one can obtain the effective cross
sections of neutrino-deuterium reactions as follows
\cite{Burrows1}\cite{Cross-Section}:
\begin{eqnarray}
&&\
\langle\sigma(d(\nu_e,e^{-})pp)\rangle=(3.35T_{\nu_e}^{2.31}-3.70)\times10^{-43}
cm^2, \nonumber \\
&&\
 \langle\sigma(d(\bar{\nu}_e,e^{+})nn)\rangle=(3.05T_{\bar{\nu}_e}^{2.08}-7.82)
 \times10^{-43} cm^2,  \label{sigma DB}
\end{eqnarray}
for the charged-current capture, and
\begin{eqnarray}
 &&\ \langle\sigma(d(\nu_{\alpha},\nu_{\alpha})np)\rangle =
  (1.63T_{\nu_{\alpha}}^{2.26}-2.78)\times10^{-43} cm^2, \nonumber \\
 &&\ \langle\sigma(d(\bar{\nu}_{\alpha},\bar{\nu}_{\alpha})np)\rangle=
 (2.03T_{\bar{\nu}_{\alpha}}^{2.05}-3.76)\times10^{-43}cm^2,  \label{sigma DC}
\end{eqnarray}
for the neutral-current capture, where $\alpha=e, ~\mu, ~\tau$.

In Figs. 9a and 9b we plot $R^+_1$ and $R^-_1$ of the
neutrino-deuterium reactions in analogy with Figs. 8a and 8b.
Similar to the neutrino-carbon case given in Fig. 8a, Fig. 9a for
$R^+_1$ can be used to determine the neutrino mass hierarchy. But it
is difficult to pin down the value of $\theta_{13}$.  Figure 9b for
$R^-_1$ is similar to Fig. 7 and the discussion can be applied to
the present situation, with the allowed value of $R^-_1$ between 1.9
to 2.6 in the case of inverted mass hierarchy.

In the above discussions, we have applied our method to three
reactions: inverse beta decay, neutrino-carbon, and
neutrino-deuterium. A summary is given in Table II. From this table,
it can be seen that the mass hierarchy may be identified by
measuring $R^+_1$ in the neutrino-carbon reactions and the
neutrino-deuterium reactions, and by measuring $R^-_1$ in the
inverse beta decay, the neutrino-carbon reactions, and the
neutrino-deuterium reactions. In addition, in the case of normal
mass hierarchy, $R^+_1$ measured in the channel of neutrino-carbon
reactions can provide the most accurate information about
$\theta_{13}$. Similarly, the present method can also be applied to
other neutrino experiments, such as KamLAND \cite{KamLAND}, MinBooNE
\cite{MinBooNE}, Borexino \cite{Borexino}, and Double-Chooz
\cite{Chooz} in the channel of neutrino-carbon reactions. In the
future, if a SN explosion takes place within the cosmic distance
considered here, it is possible to identify the mass hierarchy and
reach a much smaller value of $\theta_{13}$ than current neutrino
experiments through the measurement of $R^+_1$ in the channel of
neutrino-carbon reactions.

\section{\label{sec:cpv1}Check Neutrino Masses under cosmic
setting}

All of our understanding of neutrino properties has been derived
in the terrestrial environment, with input from solar neutrinos,
and our study of SN neutrinos has been made under the reasonable
assumption that the terrestrial properties can be applied
unaltered to the extreme conditions of the cosmos, such as those
in the SN, e.g., high temperature, high density, etc.. If an
observation of SN neutrinos deviates from the prediction, it will
be argued that our picture of SN requires modification. However,
it would be prudent to check if some aspects of the properties of
neutrinos may have been modified under the unusual cosmic
conditions. In this section we propose a measurement, the
observation of which, can only be attributed to some change of the
neutrino property, albeit a very specific one.

Since neutrinos, at least some of the species, have finite masses,
their transmitting velocities are smaller than that of light. Then,
neutrinos from a SN explosion, while traveling through a large
distance to the Earth, can experience a delay in the arrival time,
which depends on their energies and masses and may be observable
\cite{Beacom1}\cite{Beacom2}\cite{Beacom3}.

Suppose a neutrino from a SN, of mass $m$ (in eV) and energy $E$
(in MeV), travels over a distance $D$ (in 10kpc) to the Earth, it
will experience an energy-dependent time delay of $\Delta t$ (in
second) relative to a massless neutrino \cite{Beacom1}:
\begin{eqnarray}
\Delta t(s)\backsimeq 0.5143 \Bigg(\frac{m(eV)}{E(MeV)}\Bigg)^2
D(10kpc), \label{DT}
\end{eqnarray}
where only the lowest order term in the neutrino mass expansion has
been kept. In Ref. \cite{Beacom1}, a method of measuring the
neutrino mass is proposed, using the detection of the event number
of the delayed SN neutrinos. In the papers, a simple model of the
luminosities of SN neutrinos was assumed, in the absence of various
effects considered in recent SN neutrino studies. The neutrino
luminosities were assumed to have a common step-function cutoff at a
fixed time. This form of luminosities was derived from the argument
that the neutrino gravitational redshifts become severe enough only
near the sharp cutoff time $t_{BH}$, where $t_{BH}$ can be
determined from the observed time profile of SN neutrinos.
Therefore, the luminosities, which are constant before $t_{BH}$ and
vanishes abruptly afterward, take the form
$L(t)=L^0_{\alpha}\theta(t_{BH}-t)$. In the present work, we take
into account the full complexity of the neutrino flavor
transformation due to the effects of the shock, MSW, collective, and
the Earth matter. Also, we consider the time-dependence of the SN
neutrino luminosities given in Eq. (\ref{L}) and the uncertainties
in the SN neutrino flux parameters given in Eqs. (\ref{L0}) and
(\ref{T}). Since the time delay $\Delta t$ depends on the neutrino
mass $m$, we can express the event number of the delayed SN
neutrinos as a function of $m$. Therefore, information about $m$ can
be obtained by detecting the event number of the delayed SN
neutrinos.

From Ref. \cite{PDG} there are the following neutrino mass
bounds\footnote{Note that these bounds are obtained from direct
measurements of neutrino masses and they do not take into account
neutrino oscillation experiments. The latter will relate neutrino
effective masses of different flavors in a very tight range.}
\begin{eqnarray}
m_{\bar{\nu}_e}<2eV, \quad m_{\nu_{\mu}}<0.19MeV, \quad
m_{\nu_{\tau}}<18.2MeV. 
\label{neumass}
\end{eqnarray}
For the SN electron antineutrino the averaged energy is $\langle
E_{\bar{\nu}_e}\rangle\simeq16MeV$, which together with $D=10kpc$,
gives $\Delta t_{\bar{\nu}_e}<0.01s$ from Eq. (\ref{DT}). This
time delay of the $\bar{\nu}_e$ is much smaller than the time
scale of the SN explosion of about $10s$. And it is also too short
to be be observable. Similarly, for the electron neutrino
$\langle E_{\nu_e}\rangle\simeq11MeV$, 
we find $\Delta t_{\nu_e}<0.02s$, which is again very short and
much smaller than the time scale of SN explosion. Therefore,
unless their effective mass is somehow significantly enhanced,
$\nu_e$ and $\bar{\nu}_e$ can be considered as massless for the
time delay observation. For the muon and tau neutrinos, according
to the oscillation experiments their effective masses in the
terrestrial setting will also be bound by $2eV$ because of the
much smaller values of $\Delta{m}^2_{21}$ and
$\Delta{m}^2_{31}$,\footnote{The effective masses of the flavor
neutrinos, $m_\alpha$, $\alpha = e$, $\mu$, and $\tau$, to those
of the regular neutrino masses, $m_j$, $j=1,~2$, and 3, by
\begin{equation}
m_\alpha = \sum_{j=1,2,3}|U_{\alpha j}|^2 m_j, \nonumber
\end{equation}
where $U_{\alpha j}$ is the neutrino mixing matrix.} their time
delay will be very small too. In the following, we will
tentatively ignore these terrestrial bounds to form the basis of
our proposal for a test of neutrino properties under the cosmic
setting of extreme conditions.

After the SN explosion the emission of different flavors of
neutrinos lasts for a period of about $10s$. The arrival to the
Earth of those flavors of neutrinos and antineutrinos whose
effective masses are not affected will last about the same period
of time. For other flavors which are affected there may be a delay
in the arrival time, depending on the values of their effective
masses. We look for measurements which are sensitive to both
charge and neutral current events, such as the detector types of
the Daya Bay \cite{Guo2}, LVD \cite{LVD}, and SNO \cite{SNO2}
experiments, which involve the neutrino-carbon and
neutrino-deuterium reactions. The experimental information that
these detectors can potentially provide include the event numbers
of the electron antineutrino from charged-current reactions and
the total event number of all the three flavors of neutrinos from
the neutral-current reactions. We shall assume that the electron
neutrino mass is not affected so as to avoid any possibility of
influencing the existing result of nucleosynthesis.  But the
effective masses of the muon and tau neutrinos might be affected.
Then, there might be a time delay $\Delta t_{x}$ between the
charged-current events and some of the neutral-current events if
the muon and tau flavors of SN neutrinos appear to have
sufficiently sizable effective  masses. Denoting the effective
mass of $\nu_x$ ($\nu_{\mu}$, $\bar{\nu}_{\mu}$, $\nu_{\tau}$, or
$\bar{\nu}_{\tau}$) by $m_{\nu_x}$, we have
\begin{eqnarray}
\Delta t_{x}=\langle t\rangle_{NC}-\langle t\rangle_{CC}\backsimeq
0.5143 \Bigg(\frac{m_x}{E}\Bigg)^2 D, \label{DTx}
\end{eqnarray}
where we have dropped the units for $\Delta{t}_x$, $m_x$, $E$, and
$D$ as indicated in Eq. (\ref{DT}).  $<t>_{NC}$ is the average
time of SN neutrinos of the neutral-current reactions to arrive at
the detector and $<t>_{CC}$ that of the charged-current reactions.
A nonvanishing $\Delta{t}_x$ signalizes a significant larger
effective mass for muon or tau neutrino, independent of parameters
of the SN and neutrinos.

To gain a sense of how $\Delta{t}_x$ may look like and how
sensitive it is to the effective mass of the muon and tau
neutrinos, we have to do some model calculation for quantities
which can be measured in a SN neutrino experiment. We found that
such a quantity is available as described below. We use the
charged-current events which are from the electron antineutrino to
monitor the events that are not delayed, and the neutral-current
events which involve all flavors to measure the possible time
delay.  We divide the total neutral-current events into two
groups: $N^{NC(r)}_{total}$ and $N^{NC(0)}_{total}$, where
$N^{NC(r)}_{total}$ is the total neutral-current events in
accompanying with the charged-current events and are therefore not
delayed, and $N^{NC(0)}_{total}$ are all the total neutral-current
events including both delayed and not delayed. We define the event
number of the delayed $\nu_x$ plus $\bar{\nu}_x$ from the
neutral-current events as
\begin{eqnarray}
N^{NC}(delay)=N^{NC(0)}_{total}-N^{NC(r)}_{total}. \nonumber
\end{eqnarray}
The quantity to be measured is the new ratio,
\begin{eqnarray}
 R_2&=&\frac{N^{NC}(delay)}{N^{NC(0)}_{total}}
 =\frac{N^{NC(0)}_{total}-N^{NC(r)}_{total}}
  {N^{NC(0)}_{total}}. \label{ratio2}
\end{eqnarray}
This ratio is very insensitive to the values of $\Delta{m}^2_{21}$
and $\Delta{m}^2_{31}$.\footnote{Our numerical simulation shows that
$R_2$ is insensitive to variations of $\Delta{m}^2_{21}$ and
$\Delta{m}^2_{31}$ in a range of 10 and 8 orders of magnitude
respectively.} It should also be remarked that because of the ratio,
the effect of the various uncertainties involved, such as the
neutrino and SN parameters, will be reduced in $R_2$. Using the
charged current events to monitor the undelayed neutral current
events will bypass the question of the uncertainty in the total
emission time of SN neutrinos.

Since we can not tag the flavor of neutral-current events the
information on the masses of the muon and tau neutrinos are
correlated.  For the purpose of illustration we assume $\nu_{\mu}$
and $\nu_{\tau}$ to have the same effective mass.  We keep all other
features of neutrinos and the SN unchanged.  For the neutrino-carbon
reactions at Daya Bay, we plot in Fig. 10 two sets of curves which
describe $R_2$ as a function of $m_{\nu_x}$, each for a fixed value
of $\theta_{13}$. In Fig. 10(a) for $\theta_{13}=0$, it can be seen
that the curves are insensitive to $L^0_{\nu_e}/L^0_{\nu_x}$ and the
mass hierarchy. For $m_{\nu_x}<200eV$, $R_2$ is sensitive to
$m_{\nu_x}$.  A measurement of $R_2$ can restrict $m_x$ in a
distinctive range of values determined by the uncertainties in the
SN neutrino temperature and chemical potential. In Fig. 10(b) for
$\theta_{13}=3^\circ$, $R_2$ is again shown to be sensitive to $m_x$
for $m_{\nu_x}<200eV$. The two mass hierarchies overlap for small
$m_x$ but fall into well separated regions for $m_x>200eV$.

In Figs. 11 we plot $R_2$ for the neutrino-deuterium reactions in
analogy to Figs. 11 of the neutrino-carbon reactions. Figure 11(a)
is very similar to Fig. 10(a). Figure 11(b) is also similar to Fig.
10(b) for $m_{\nu_x}$ smaller than $150eV$, but there are some
differences at larger $m_x$.  In this latter case, unlike Fig.
10(b), Fig. 11(b) shows that the two mass hierarchies can no long be
separated even for $m_{\nu_x}$ greater than $200eV$.

The method can also be applied to KamLAMD, MinBooNE, Borexino, and
Double-Chooz in the channel of neutrino-carbon reactions, and the
results are similar to those given for the Daya Bay experiment.

\section{\label{sec:cpv1}Discussion and Summary }

In this paper, we considered the SN shock effects, the MSW effects,
the collective effects, and the Earth matter effects in the
detection of type II SN neutrinos.  Also, we considered the
uncertainties in the neutrino luminosities $L^0_{\alpha}$, their
temperatures $T_{\alpha}$, and their pinching parameters
$\eta_{\alpha}$ in the calculation of different flavors of SN
neutrinos.  We found that quantities suitably defined in terms of
event numbers of different flavors of SN neutrinos are sensitive to
the neutrino mass hierarchy, the mixing angle $\theta_{13}$, and
neutrino masses. Therefore, it is possible to identify the mass
hierarchy, acquire information about $\theta_{13}$, and look into
certain neutrino mass patterns of SN neutrinos.

Firstly, we defined the ratio $R^{+(-)}_1$ as the event number of
$\nu_e$($\bar{\nu}_e$) for the charged-current reaction detected
after one second over that detected in the first one second, and
discussed the dependence of $R^{+(-)}_1$ on $\theta_{13}$ for
different mass hierarchies. With this, we may identify the mass
hierarchy and obtain information about $\theta_{13}$. This method
can be applied to several detector types relying on specific
physical processes, including the inverse beta decay (Fig. 7),
neutrino-carbon reactions (Figs. 8(a) and 8(b)), and
neutrino-deuterium reactions (Figs. 9(a) and 9(b)). From our
numerical calculations, the mass hierarchy may be identified by
measuring $R^+_1$ in the neutrino-carbon reactions and the
neutrino-deuterium reactions, $R^-_1$ in the inverse beta decay, the
neutrino-carbon reaction, and the neutrino-deuterium reactions. In
addition, in the case of normal mass hierarchy, $R^+_1$ measured in
the neutrino-carbon reactions provides the most restrictive
information about $\theta_{13}$. Therefore, we can both identify the
mass hierarchy and measure $\theta_{13}$ from $R^+_1$ in the channel
of neutrino-carbon reactions at several experiments such as Daya
Bay, KamLAND, MinBooNE, Borexino, and Double-Chooz.

Secondly, we defined another ratio, $R_2$, for the neutral-current
reactions, as the event number of the delayed neutrinos over the
total event number, with the charge current events as a monitor of
undelayed events.  The significance of $R_2$ is that it offers a
possibility to detect possible difference between the electron
neutrino and its other flavors counterparts, in particular, any
significant difference in their effective masses under the cosmic
setting of extreme conditions.  With the plot of $R_2$ vs
$m_{\nu_x}$, we can study $m_{\nu_x}$ vs $m_{\nu_e}$. This method is
applied to both neutrino-carbon and neutrino-deuterium reactions and
the result is given respectively in Figs. 10 and 11. What we have
presented for $R_2$ is a case study with rather restricted input. In
case there is an indication of non-vanishing $R_2$ a detailed
refined calculation should be made.

In conclusion, if a SN explosion takes place within the cosmic
distance considered here, it is possible to identify the mass
hierarchy and obtain information on $\theta_{13}$. With suitable
set up of the detector, a study of some unusual properties of
neutrinos under the cosmic setting is possible. A summary of the
detector types and their capabilities in association with the
studies presented in this article is given in Table II.

\section{Acknowledgments}

We would like to thank F.-G. Cao and C.-G. Yang for helpful
discussions. This work was supported in part by National Natural
Science Foundation of China (Project Numbers 10535050, 10675022,
10975018, and 10890092) and the Special Grants from Beijing Normal
University.


\newpage

\noindent{\large \bf Figure Captions} \\
\vspace{0.4cm}

\noindent Fig. 1 The density of the SN as a function of the radius
$r$ at different times.\vspace{0.2cm}

\noindent Fig. 2 The crossing probability at the high resonance
region $P_H$ as a function of the neutrino energy $E$ for two very
small neutrino mixing angles $\theta_{13}$ at $t=4s$.
\vspace{0.2cm}

\noindent Fig. 3 The crossing probability at the high resonance
region $P_H$ as a function of time $t$ for three different
neutrino energies at $\sin^2\theta_{13}=10^{-4}$. \vspace{0.2cm}

\noindent Fig. 4 The crossing probability at the high resonance
region $P_H$ as a function of the neutrino mixing angle
$\theta_{13}$: (a) for three different times at $E=11MeV$. The
solid, dashed, and dotted curves are for $t=2s, 4s, 6s$,
respectively; (b) for three different neutrino energies at $t=4s$.
The solid, dashed, and dotted curves are for $E=11MeV, 16MeV,
25MeV$, respectively. \vspace{0.2cm}

\noindent Fig. 5 Illustration of the path of the SN neutrino
reaching the detector in the Earth. $D$ is the location of the
detector, $\theta$ is the incident angle of the neutrino,
 $O$ is the center of the Earth, $L$ is the distance the neutrino
travels through the Earth matter, and $\tilde{x}$ is the
instantaneous distance of the neutrino to the center of the Earth.
\vspace{0.2cm}

\noindent Fig. 6  The luminosity decay time $\tau$ as a function of
the mixing angle $\theta_{13}$ in the channel of neutrino-carbon
reactions at Daya Bay when $R^+_1=2.4$ and
$L^0_{\nu_e}/L^0_{\nu_x}=1$. The "max" ("min") corresponds to the
maximum (minimum) values of $T_{\alpha}$ and $\eta_{\alpha}$.
\vspace{0.2cm}

\noindent Fig. 7 The ratio $R^-_1$ of the event number of
$\bar{\nu}_e$ for the charged-current reaction detected after 1s
over that detected before 1s, as a function of the mixing angle
$\theta_{13}$ in the channel of inverse beta decay at Daya Bay. The
solid curves correspond to $L^0_{\nu_e}/L^0_{\nu_x}=1/2$, the dashed
curves $L^0_{\nu_e}/L^0_{\nu_x}=1$, and the dot-dashed curves
$L^0_{\nu_e}/L^0_{\nu_x}=2$. The "max" ("min") corresponds to the
maximum (minimum) values of $T_{\alpha}$ and $\eta_{\alpha}$.
\vspace{0.2cm}

\noindent Fig. 8 (a) Similar to Fig. 7 but the electron neutrinos
$\nu_e$ are observed in the channel of neutrino-carbon reactions at
Daya Bay; (b) similar to Fig. 7 but the anti-electron neutrinos
$\bar{\nu}_e$ are observed in the channel of neutrino-carbon
reactions at Daya Bay. \vspace{0.2cm}

\noindent Fig. 9 (a) Similar to Fig. 7 but the electron neutrinos
$\nu_e$ are observed in the channel of neutrino-deuterium reactions
at SNO; (b) similar to Fig. 7 but the anti-electron neutrinos
$\bar{\nu}_e$ are observed in the channel of neutrino-deuterium
reactions at SNO. \vspace{0.2cm}

\noindent Fig. 10 For the neutral-current reactions $R_2$ in the
channel of neutrino-carbon reactions at Daya Bay, the ratio of the
event number of delayed neutrinos over the total event number, as a
function of $m_{\nu_x}$ at two values of $\theta_{13}$, each has the
two mass hierarchies with $L^0_{\nu_e}/L^0_{\nu_x}$, $T_{\alpha}$,
and $\eta_{\alpha}$ taking their limiting values.  The solid curves
correspond to $L^0_{\nu_e}/L^0_{\nu_x}=1/2$, the dashed curves
$L^0_{\nu_e}/L^0_{\nu_x}=1$, and the dot-dashed curves
$L^0_{\nu_e}/L^0_{\nu_x}=2$. The "max" ("min") corresponds to the
maximum (minimum) values of $T_{\alpha}$ and $\eta_{\alpha}$. Figure
(a) shows that $R_2$ is nearly independent of
$L^0_{\nu_e}/L^0_{\nu_x}$ and the mass hierarchy. \vspace{0.2cm}

\noindent Fig. 11 Similar to Fig. 10 but in the channel of
neutrino-deuterium reactions at SNO.

\newpage

\begin{figure}
\includegraphics[width=0.4\textwidth]{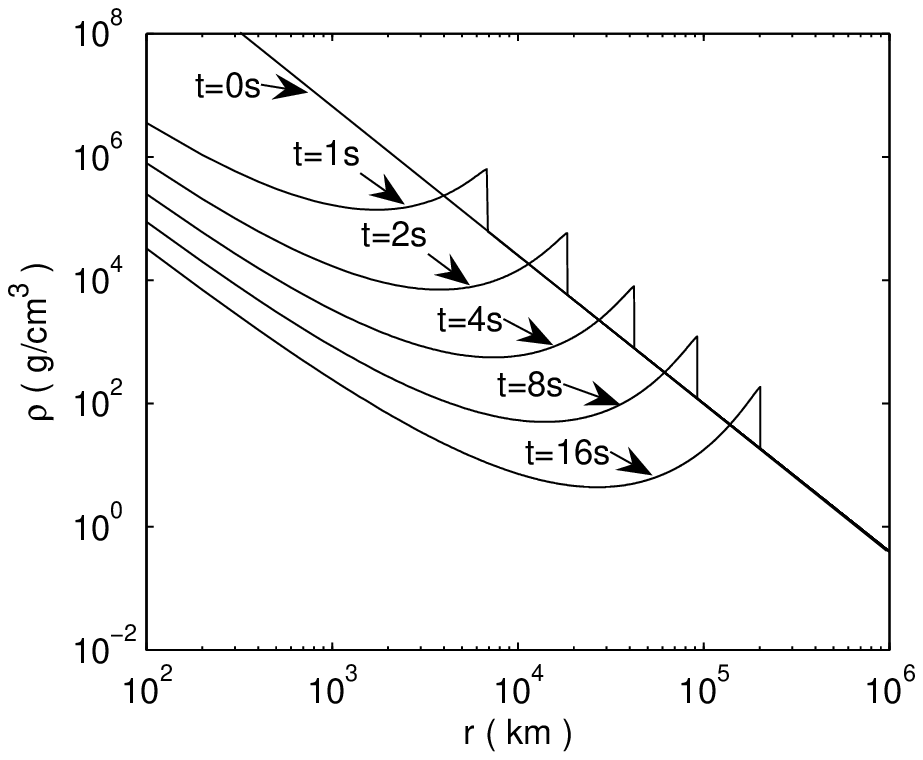}\\
\centerline{Fig. 1}
\end{figure}

\begin{figure}
\includegraphics[width=0.7\textwidth]{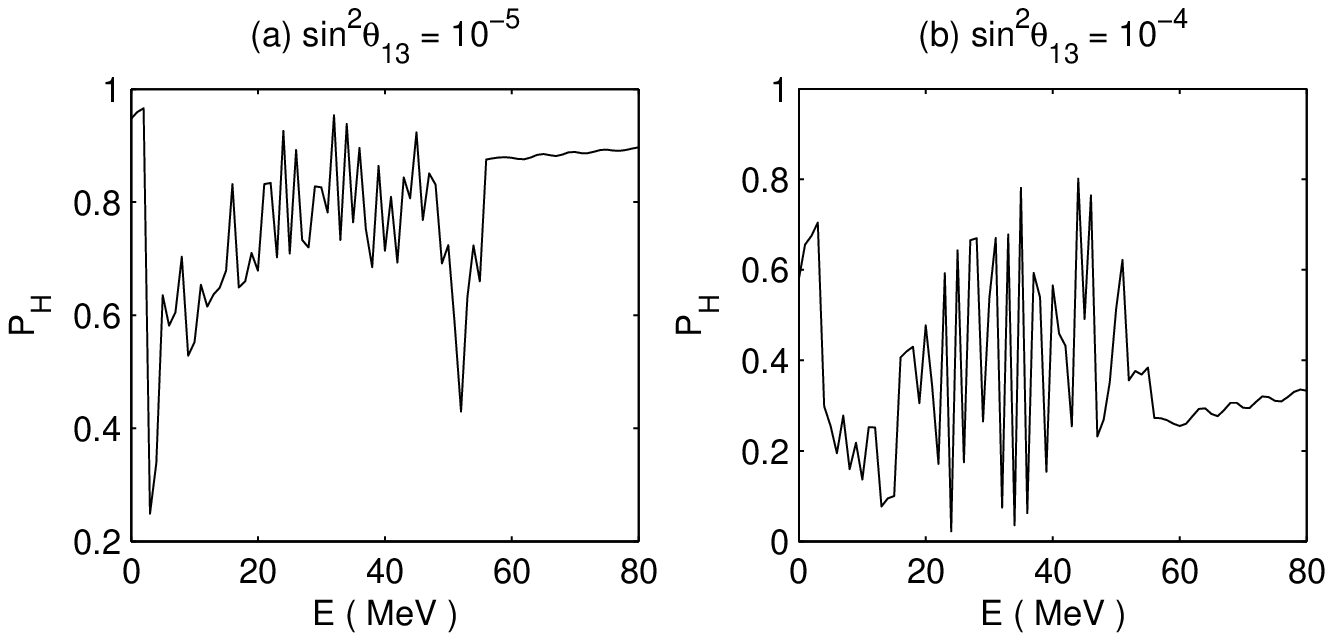}\\
\centerline{Fig. 2}
\end{figure}

\begin{figure}
\includegraphics[width=1\textwidth]{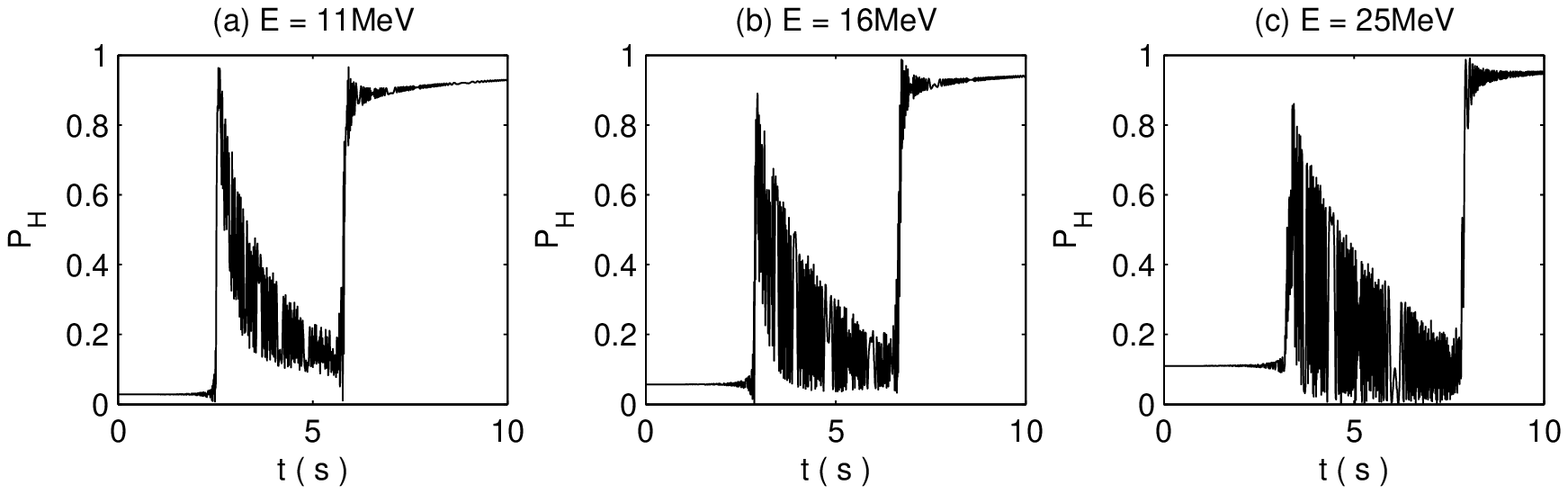}\\
\centerline{Fig. 3}
\end{figure}

\begin{figure}
\includegraphics[width=0.7\textwidth]{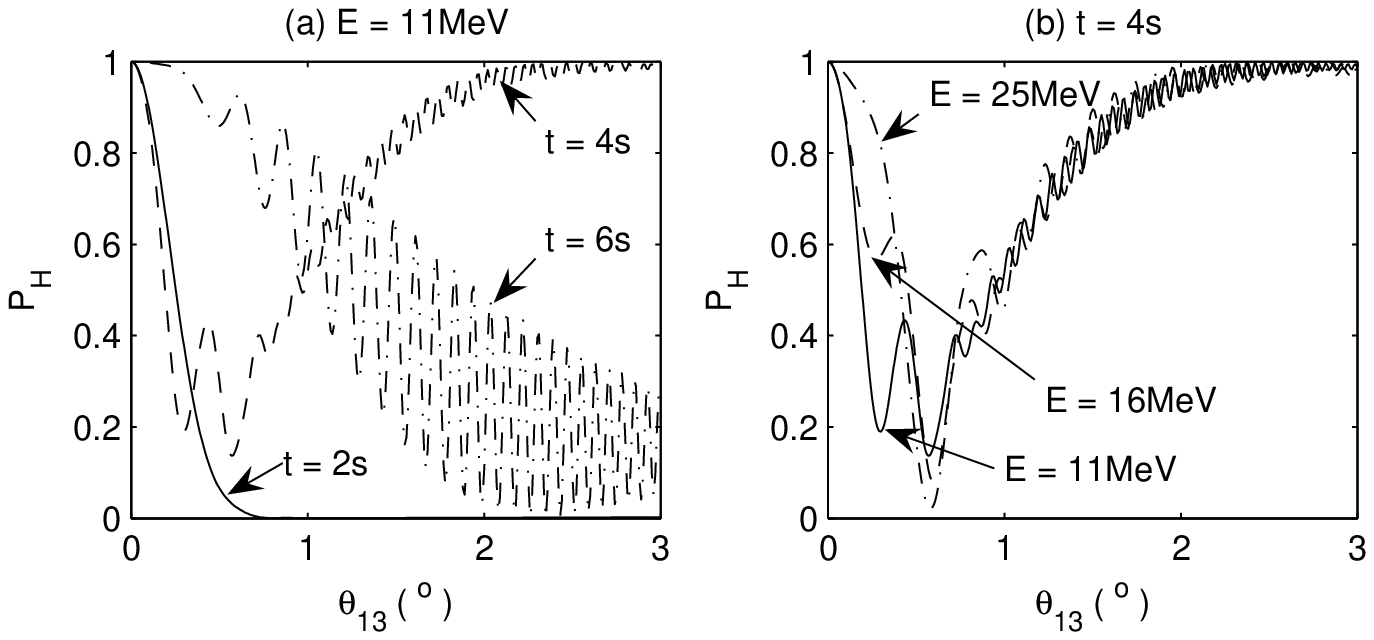}\\
\centerline{Fig. 4}
\end{figure}

\begin{figure}
\includegraphics[width=0.35\textwidth]{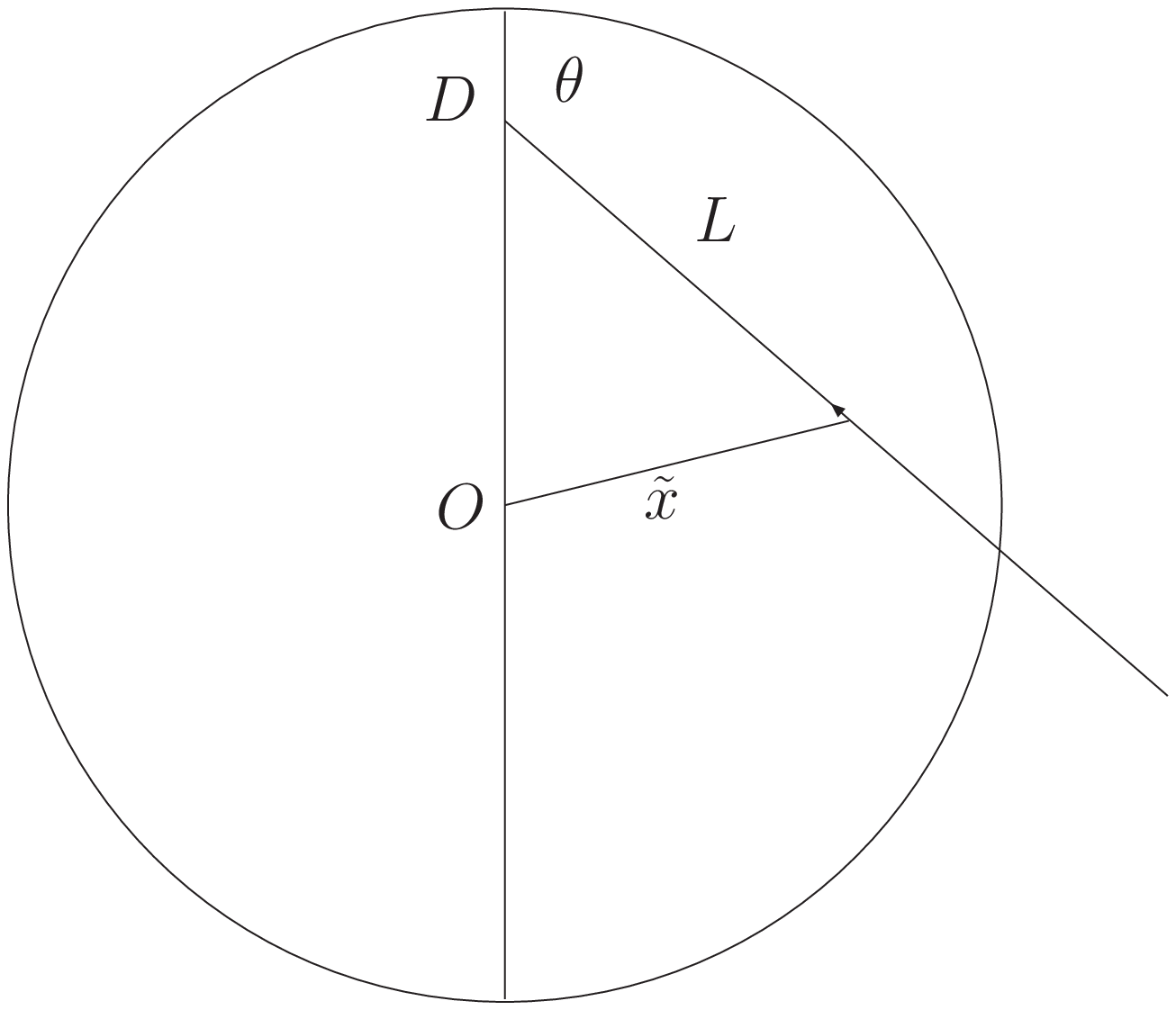}\\
\centerline{Fig. 5}
\end{figure}

\begin{figure}
\includegraphics[width=0.4\textwidth]{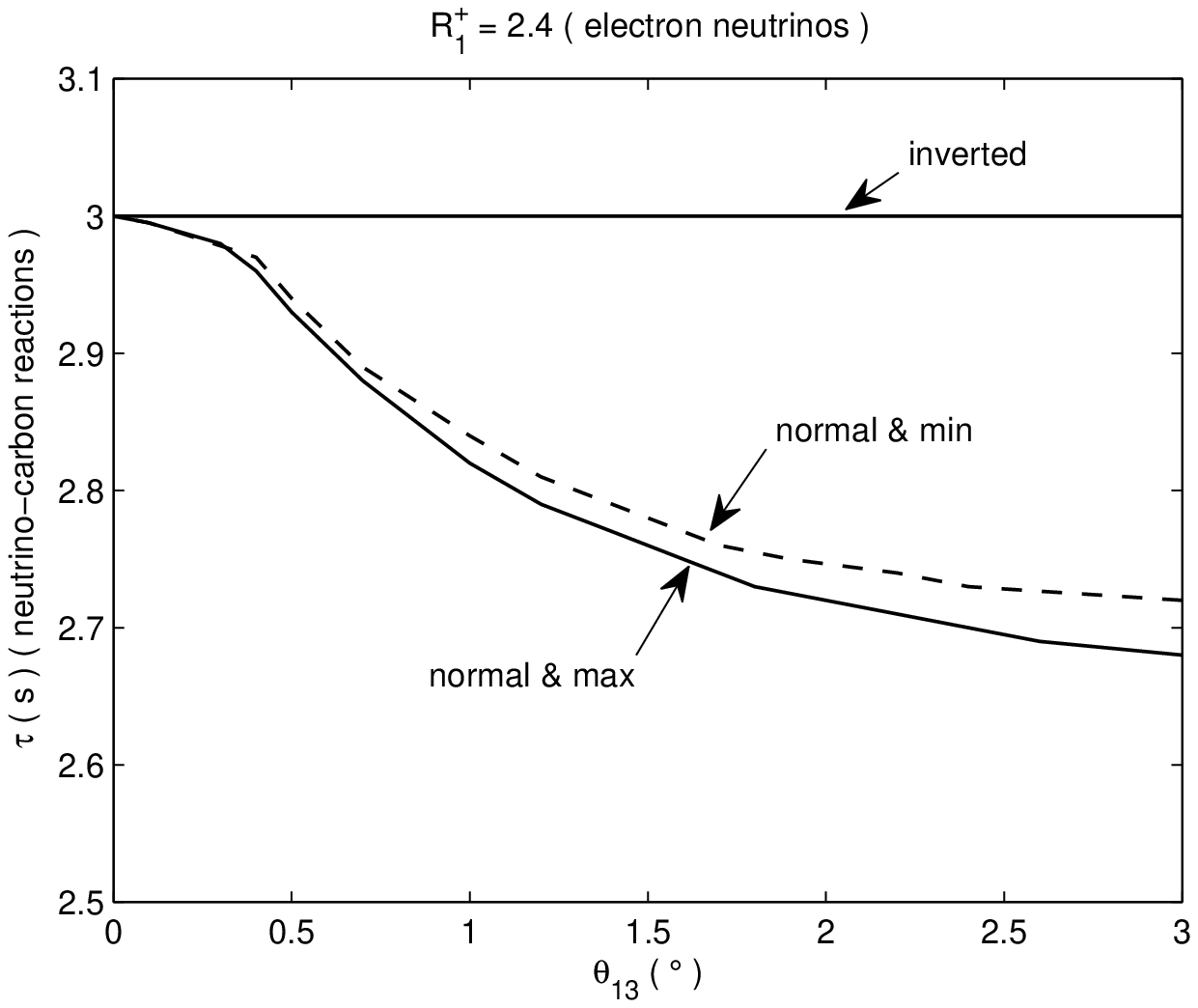}\\
\centerline{Fig. 6}
\end{figure}

\begin{figure}
\includegraphics[width=0.4\textwidth]{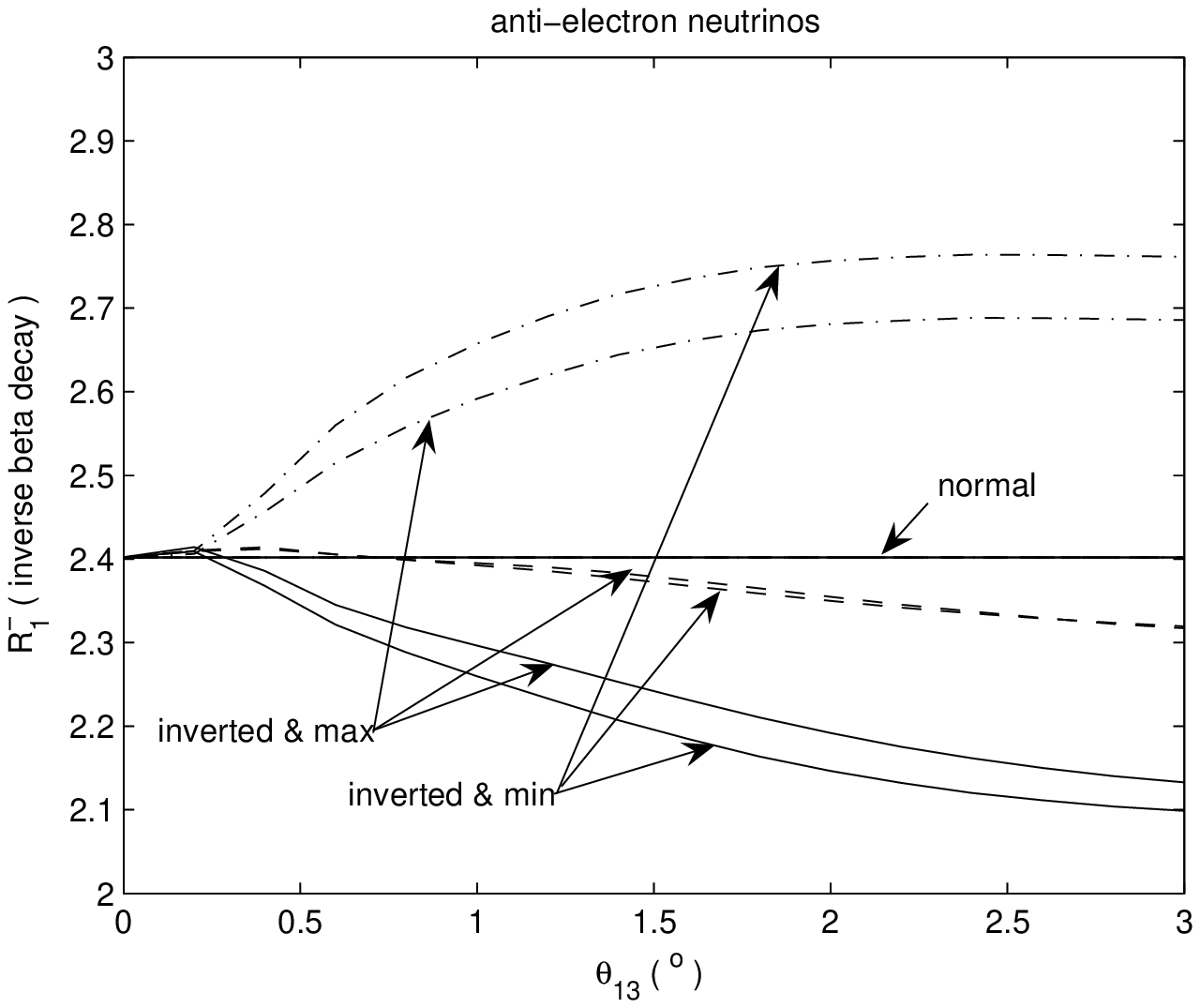}\\
\centerline{Fig. 7}
\end{figure}

\begin{figure}
\includegraphics[width=0.38\textwidth]{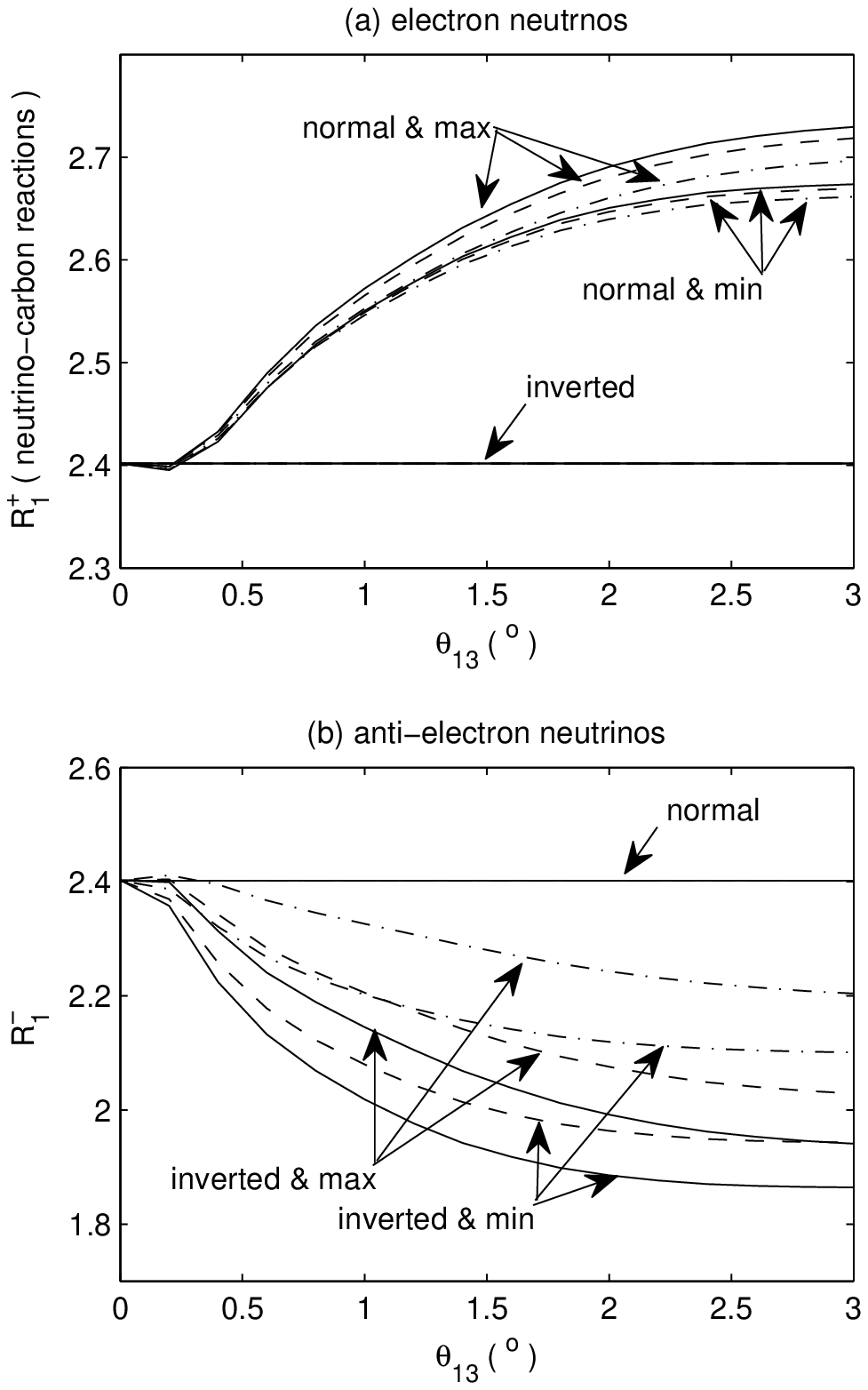}\\
\centerline{Fig. 8}
\end{figure}

\begin{figure}
\includegraphics[width=0.38\textwidth]{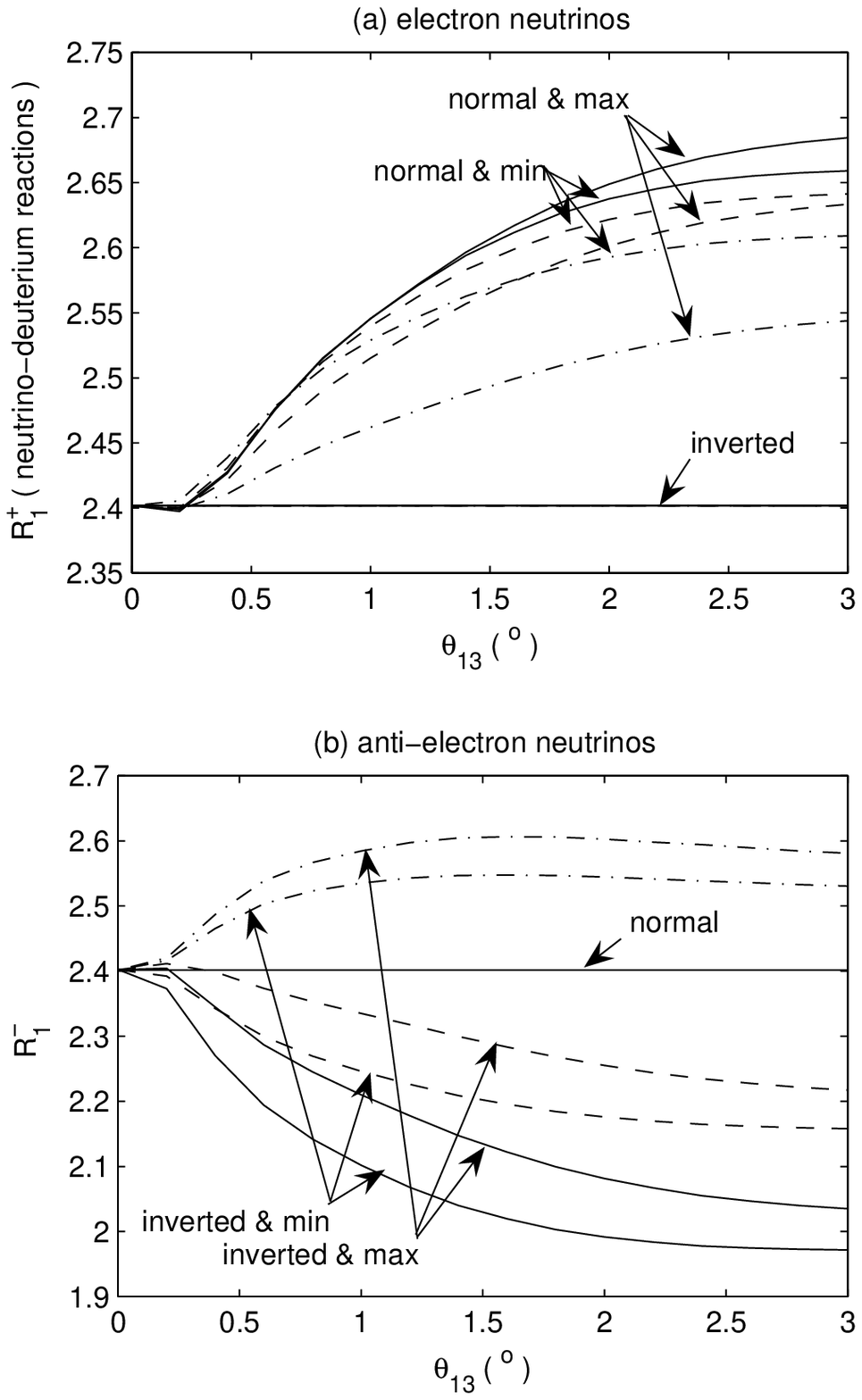}\\
\centerline{Fig. 9}
\end{figure}

\begin{figure}
\includegraphics[width=0.38\textwidth]{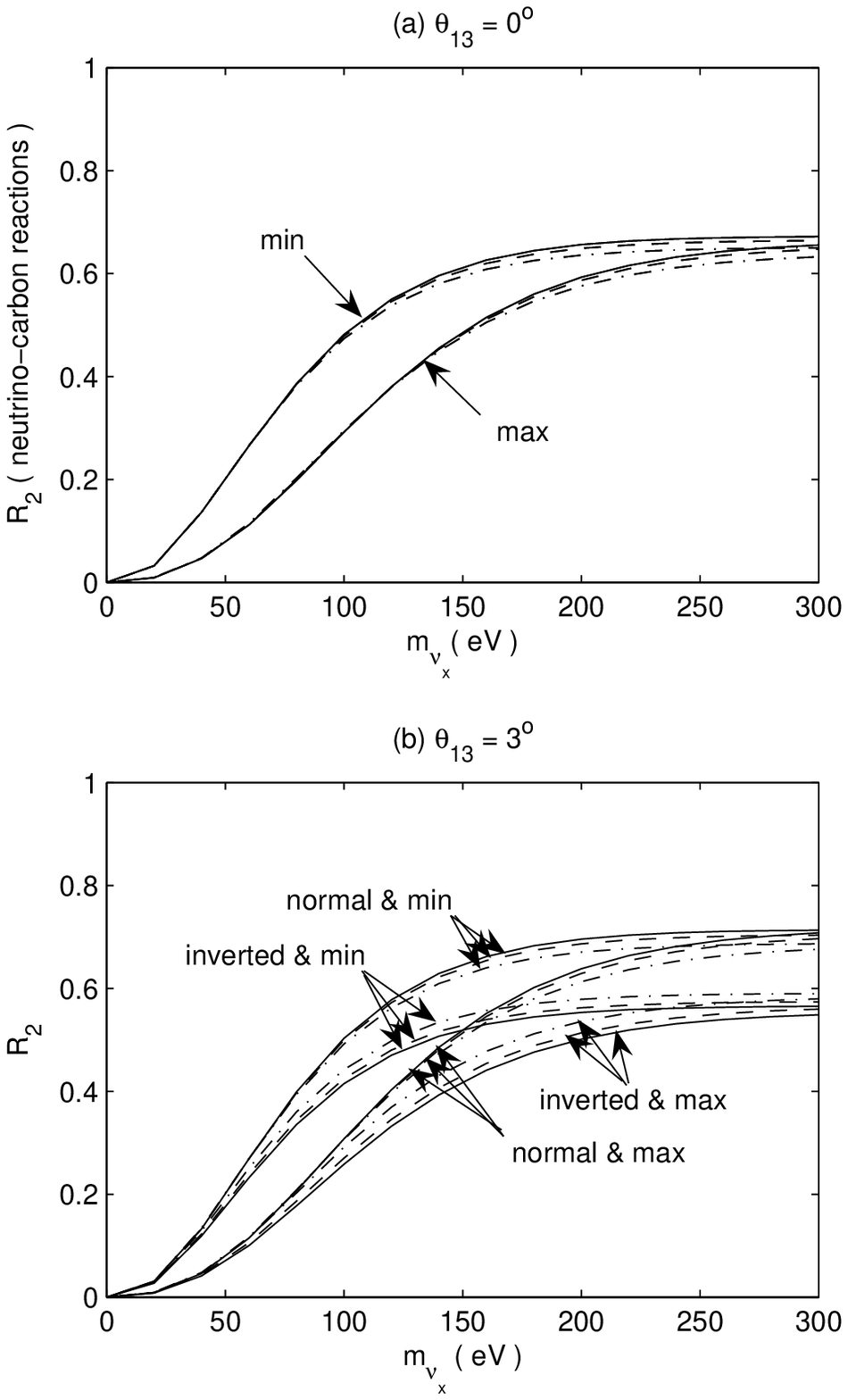}\\
\centerline{Fig. 10}
\end{figure}

\begin{figure}
\includegraphics[width=0.38\textwidth]{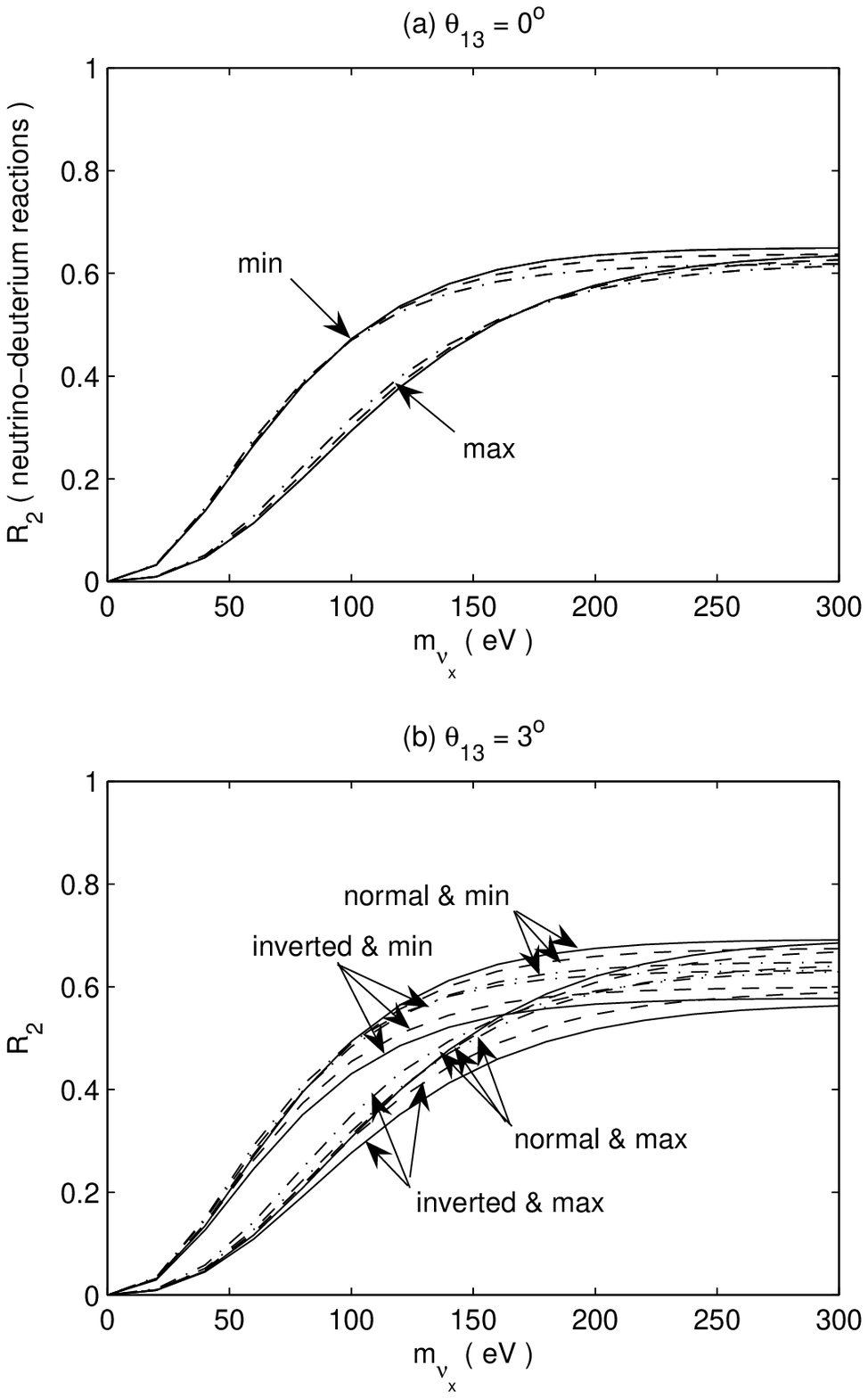}\\
\centerline{Fig. 11}
\end{figure}

\cleardoublepage

\begin{table}[hbtp!]
\begin{center}
\caption{Summary of neutrino flavor conversions due to the neutrino
shock effects and the MSW effects in various density regions.}
\label{tab:Probab} \vspace{0.3cm}
\begin{tabular}{|c||c|c||c|c|}\hline
$t<1s$ & \multicolumn{4}{c|}{$t\geq 1s$} \\ \hline\hline
    & \multicolumn{2}{c||}{$\rho_b>\rho_-$}
     & \multicolumn{2}{c|}{$\rho_-\geq\rho_b$}
                \\ \cline{2-5}  
  & resonance in & flavor conversion
     & Resonance in & flavor conversion  \\
  & region & involved & region  & involved
                 \\ \cline{2-5}
  $P_{H3}$ & $\rho_{res}>\rho_+$ & $P_{H1},~P_s$
     & $\rho_{res}>\rho_+$ & $P_{H1},~P_s$
                 \\ \cline{2-5} 
  & $\rho_+\geq\rho_{res}>\rho_b$ & $P_{H1},~P_{H2},~P_s$
       & $\rho_+\geq\rho_{res}>\rho_-$ & $P_{H1},~P_{H2},~P_s$
                 \\ \cline{2-5} 
  & $\rho_{res}=\rho_b$ & $P_{H1},~P_s$
       & $\rho_-\geq\rho_{res}>\rho_b$ & $P_{H1},~P_{H2},~P_s,~P_{H3}$
                 \\ \cline{2-5} 
  & $\rho_b>\rho_{res}>\rho_-$ & $P_s$
       & $\rho_{res}=\rho_b$ & $P_{H1},~P_s~P_{H3}$ \\ \cline{2-5}
  & $\rho_-\geq\rho_{res}$ & $P_s,~P_{H3}$
     & $\rho_b>\rho_{res}$ & $P_s,~P_{H3}$ \\ \hline
\end{tabular}
\end{center}
\end{table}

\begin{table}[!htb]
\begin{ruledtabular}
\caption{Summary of measurement of the neutrino parameters by SN
neutrinos in three reactions: the inverse beta decay,
neutrino-carbon reactions, neutrino-deuterium reactions. "$\surd$"
and "$\times$" represents respectively the information can and can
not be
obtained. 
} \vspace{0.3cm}
\begin{tabular}{|l|ll|l|l|l|}
  Reaction & normal hierarchy  & inverted hierarchy  & $\theta_{13}$ ($\leq3^o$)
     & cosmic setting & figures \\
  {}  & ($\Delta m^2_{31}>0$)  & ($\Delta m^2_{31}<0$) & {}
 & ($m_{\nu_{\mu}}$, $m_{\bar{\nu}_{\mu}}$, $m_{\nu_{\tau}}$, $m_{\bar{\nu}_{\tau}}$)
 &  \\
 \hline
  $\bar{\nu}_e+p$  & $\times$  & $\surd$ ($R^-_1\neq2.4$) & $\times$ & $\times$
  & Fig. 7  \\
 $\nu_{\alpha}+C$  & $\surd$ ($R^+_1>2.4$) & $\times$ & $\surd$ & $\surd$
  & Figs. 8(a), 10(a), 10(b)\\
 $\bar{\nu}_{\alpha}+C$  & $\times$  & $\surd$ ($R^-_1<2.4$) & $\times$ & $\surd$
  & Figs. 8(b), 10(a), 10(b)\\
 $\nu_{\alpha}+d$  & $\surd$ ($R^+_1>2.4$) & $\times$ & $\times$ & $\surd$
  & Figs. 9(a), 11(a), 11(b)\\
 $\bar{\nu}_{\alpha}+d$  & $\times$ & $\surd$ ($R^-_1\neq2.4$) & $\times$ &  $\surd$
  & Figs. 9(b), 11(a), 11(b)\\
 \end{tabular}
 \end{ruledtabular}
 \end{table}

\end{document}